\documentclass[12pt]{article}
\usepackage{}
\usepackage{stmaryrd}
\usepackage{amssymb}
\usepackage{color, amsmath,amssymb, amsfonts, amstext,amsthm, latexsym}
\usepackage{amssymb, epsfig, amssymb, latexsym}
\usepackage{amsmath}
\usepackage{graphicx}
\usepackage{longtable}
\usepackage{graphicx}
\usepackage{subfigure}
\usepackage[colorlinks=false, linktocpage=true]{hyperref}

\renewcommand{\phi}{\varphi}

\theoremstyle{definition}

\theoremstyle{remark}

\title{Likelihood for transcriptions  in a genetic regulatory system under asymmetric  stable L\'evy noise \footnote {This work was partly supported by the National Science Foundation grant 1620449,  and the National Natural Science Foundation of China grants 11531006 and 11771449.}}

\author{Hui Wang \footnote{ Center for Mathematical Sciences and School of Mathematics and Statistics, Huazhong University of Science and Technology, Wuhan 430074,  China  (huiwheda@hust.edu.cn).} , Xiujun Cheng \footnote{Center for Mathematical Sciences and School of Mathematics and Statistics, Huazhong University of Science and Technology, Wuhan 430074,  China  (xiujuncheng@hust.edu.cn).} , Jinqiao Duan\footnote{Department of Applied Mathematics, Illinois Institute of Technology, Chicago, IL 60616, USA and Center for Mathematical Sciences and School of Mathematics and Statistics, Huazhong University of Science and Technology, Wuhan 430074,  China   (duan@iit.edu).} , J\"{u}rgen Kurths\footnote{Department of Physics, Humboldt University of Berlin, Newtonstrate 15, 12489 Berlin, Germany (kurths@pik-potsdam.de).},  Xiaofan Li\footnote{Department of Applied Mathematics, Illinois Institute of Technology, Chicago, IL 60616, USA  (lix@iit.edu).} }

\date{\today}

\begin{document}
\maketitle
\begin{abstract}
  This work is devoted to investigating the   evolution of concentration in  a   genetic regulation system, when the synthesis reaction rate is under additive and multiplicative asymmetric  stable L\'evy   fluctuations.  By focusing on the impact of skewness (i.e., non-symmetry) in the probability distributions of noise, we find that via examining the mean first exit time (MFET) and the first escape probability (FEP), the asymmetric  fluctuations, interacting with nonlinearity in the  system,  lead to peculiar  likelihood for  transcription. This includes, in the additive noise case, realizing higher likelihood of transcription  for larger positive skewness (i.e., asymmetry) index  $\beta$,   causing   a stochastic bifurcation   at the non-Gaussianity index value $\alpha=1$ (i.e.,  it is a separating point or line for the likelihood for transcription), and achieving a turning point at the threshold value $\beta \approx -0.5$ (i.e., beyond which the likelihood for transcription suddenly reversed for $\alpha$ values).  The stochastic bifurcation and turning point phenomena do not occur in the symmetric noise case ($\beta =0$). While in the multiplicative noise case, non-Gaussianity index value $\alpha=1$ is a separating point or line for both the mean first exit time (MFET) and the first escape probability (FEP). We also investigate the noise enhanced stability  phenomenon. Additionally, we  are able to specify the regions in the whole parameter space for the asymmetric     noise,  in which we  attain  desired    likelihood for transcription.   We have conducted a series of numerical experiments in `regulating' the likelihood of gene transcription  by tuning  asymmetric  stable L\'evy noise indexes. This work offers insights for possible ways of achieving gene regulation in experimental research.

\end{abstract}


\paragraph{}
\begin{small} {\bfseries Noise plays a crucial role in gene regulation. It is  a recent challenge to better understand how noise affects gene transcriptions and protein production.  The bursty and intermittent  transcription processes    resemble the features of  a stable L\'evy motion. As  a non-Gaussian stochastic process, a stable L\'evy motion is characterized by its skewness (i.e., asymmetry)  and non-Gaussianity indexes. In extension to former results,  we study here the impact of asymmetric stable L\'evy fluctuations on the likelihood of transcriptions in a prototypical gene regulatory system. We find that the interaction between the system's nonlinearity and  fluctuations   induces   various possibilities  for transcriptions.  We discover  certain effects of the asymmetry index and other noise parameters on the likelihood for transcriptions. Hence we are able to select   combinations of these parameters, in order  to achieve the desired likelihood for transcription.}
\end{small}
\begin{center}
---------------------------------------------------------
\end{center}

\section{Introduction}
Gene regulation is a crucial  but noisy biological process  \cite{Raser2005,Maheshri2007}. The significance of noise in  genetic networks has been recognized and studied  \cite{Swain2002,Kittisopikul2010,Bressloff2014,Gui2016,Suel2007,Turcotte2008,Friedman2006,Lin2016}. It has been shown recently that noise is vital for regime transitions in gene regulatory systems \cite{Assaf2011,Hasty2000,Choi2008,Tabor2008,Munsky2012}.  In these works, noisy   fluctuations are, however, taken to have Gaussian distributions only  \cite{Gui2016,Suel2006,Hasty2000b,Liu2004,Li2014} and are expressed in terms of Brownian motion.

 But when the fluctuations are present in certain events, such as  bursty transition events, the Gaussianity  assumption is not proper. In this case, it is more appropriate to model the random fluctuations by a non-Gaussian L\'evy motion  with heavy tails and bursting sample paths \cite{Klafter2011,Woyczynski2001,Chechkin2007,Gehao2015}.
Especially, during the regulation of gene expression, transcriptions of DNA from genes and  translations into proteins occur in a bursty,  intermittent   way \cite{Friedman2006,Lin2016,Holloway2017,Kumar2015,Dar2012,Ozbudak2002,Blake2003,Sanchez2013,Raj2009}. This intermittent manner  \cite{Raj2006,Golding2005,Bohrer2016,Muramoto2012} resembles the features of a stable L\'evy motion, which is  a non-Gaussian process with jumps.

Recent studies \cite{Xu2013,Zheng2016} have recognized that  \emph{symmetric} stable  L\'evy motion can induce switches between different  gene expression states. Note that symmetry (zero skewness) in stable L\'evy motions is a special, idealized situation \cite{Applebaum2009,Duan2015,Sato1999}. The asymmetric L\'evy noise is more general and more representative.


In this present paper,  we  examine the likelihood for  transitions from low to high concentrations (i.e., likelihood for transcriptions) in  a  genetic regulatory system under  \emph{asymmetric}  (i.e., non-symmetric)  stable L\'evy noise,   highlighting the dynamical differences with the case of symmetric noise.  To this end, we compute two deterministic quantities,  the mean first exit time (MFET) and the first escape probability (FEP).
The  MFET is the mean time scale for the system to exit   the low concentration state (i.e., the longer the exit time, the less likely for transcription), while the FEP is  the switch probability  from low concentration states to high concentration states (i.e., it is the likelihood for transcriptions).

Having a better understanding of the likelihood for transcriptions in the genetic regulatory networks, we could shed light on the mechanisms of diseases which are caused by the dysregulation of gene expressions.

This paper is organized as follows. In Section 2, we briefly describe a genetic regulation model with noisy fluctuations in the synthesis reaction rate. In Section 3, we  recall basic facts about asymmetric stable L\'{e}vy motions.  In Section 4, we  investigate  the transition phenomena under additive asymmetric L\'evy motion by numerically  computing two deterministic quantities, highlighting the differences with  the symmetric stable L\'evy noise case. In Section 5, we study the multiplicative  asymmetric L\'evy motion case.  Finally, we make some concluding remarks  in Section 6. The Appendix contains the mathematical formulation for the first mean exit time and the first escape probability, in terms of deterministic integral-differential equations.

\section{A stochastic genetic regulatory system}

 Smolen et al. \cite{Smolen1998} introduced  the following  model  for the concentration  `$x$'  of the transcription factor activator (\textcolor{blue}{TF-A})
\begin{equation} \label{eq:1}
  \dot{x} = \frac{k_f x^2}{x^2 + K_d} - k_d x + R_{bas}.
\end{equation}

\begin{figure}[!htp]
\centering\includegraphics[width=0.45\textwidth]{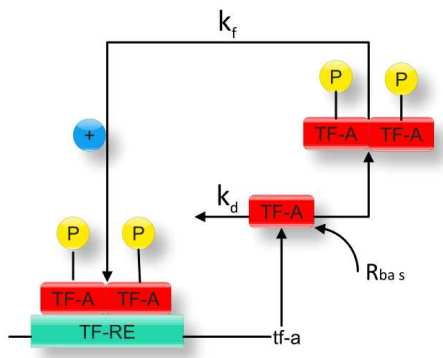}
\caption{(Color online) Genetic regulatory model with a feedforward  (Eq. \eqref{eq:1}).}
\label{Fig.1}
\end{figure}

This is a relatively   basic model of positive and negative autoregulations of transcription factors (Fig. \ref{Fig.1}). A transcription factor activator,   denoted by   (\textcolor{blue}{TF-A}),  is considered as part of a pathway mediating a cellular response to a stimulus. The transcription factor  forms a homodimer which can bind to specific responsive elements (\textcolor{blue}{TF-REs}). The \textcolor{blue}{TF-A} gene includes a \textcolor{blue}{TF-RE},  and when homodimers bind to this element, \textcolor{blue}{TF-A} transcription is increased. Only phosphorylated dimers can activate transcription. The regulatory activity of transcription factors is often modulated by phosphorylation. It is assumed that the transcription rate saturates with \textcolor{blue}{TF-A} dimer concentration to a maximal rate $k_f$,  \textcolor{blue}{TF-A} degrades with first-order kinetics with the rate  $k_d$, and \textcolor{blue}{TF-A} dimer dissociates from TF-REs with the constant $K_d$. The basal rate of the synthesis of the activator is $R_{bas}$.

With the potential
$$
U(x)= k_f\sqrt{K_d}\ \arctan\frac{x}{\sqrt{K_d}} + \frac{k_d}{2}x^2 - (R_{bas} + k_f)x,
$$
the model equation \eqref{eq:1}  becomes
$$
\dot{x} = f(x) = - U'(x).
$$
This system has two stable and one unstable equilibrium states,  i.e.,  a double-well structure,   when  the parameters satisfy the following condition:
$$
[-(\frac{k_f + R_{bas}}{3k_d})^3 + \frac{K_d(k_f + R_{bas})}{6k_d} - \frac{K_dR_{bas}}{2k_d}]^2 + [\frac{K_d}{3} - (\frac{k_f + R_{bas}}{3k_d})^2]^3 < 0.
$$

\begin{figure}[!htp]
\centering\includegraphics[width=0.45\textwidth]{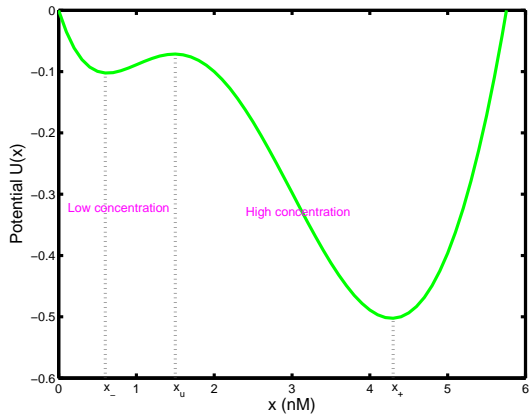}
\caption{(Color online) The bistable potential U for the  genetic regulatory model.}
\label{Fig.01}
\end{figure}
 Under  this `double-well' condition and as in Smolen et al. \cite{Smolen1998},  we choose proper parameters  in this genetic regulatory system on the basis of genetic significance and also for convenience:  $K_d = 10$, $k_d = 1 min^{-1}$,  $k_f = 6 min^{-1}$, and $R_{bas} = 0.4 min^{-1}$. Then the two stable states are
 $$x_- \approx 0.62685 nM, \;\;   x_+ \approx 4.28343 nM,  $$
 and the unstable  state (a saddle)  is
 $$x_u \approx 1.48971 nM. $$
  That is, the deterministic dynamical system   \eqref{eq:1} has two stable states: $x_-$ and $x_+$  as well as one unstable state $x_u$, See Fig. \ref{Fig.01}.

However, the basal synthesis rate $R_{bas}$ is unavoidably influenced by many   factors \cite{Smolen1998}, such as the mutations, the biochemical reactions inside the cell, and the concentration of other proteins. These fluctuations in the genetic regulatory system behaves like bursty perturbations as we discussed  in the introduction. Therefore, we incorporate an asymmetric  stable L\'evy motion as a random perturbation of the synthesis rate $R_{bas}$. Thus  the model   \eqref{eq:1}    becomes  the following stochastic gene regulation model:
\begin{equation}  \label{eq:2}
  \dot{X_t} = \frac{k_f X_t^2}{X_t^2 + K_d} - k_d X_t + R_{bas}+\dot{L}_t^{\alpha, \beta} , \qquad  X_0 = x,
\end{equation}

The effect of multiplicative noise has been investigated in the literature \cite{AM1991,CS1993,AL2010}. Here we also consider that the  synthesis rate $R_{bas}$ is perturbed by a multiplicative asymmetric  L\'evy motion as follows:
\begin{equation}  \label{multimodel}
  \dot{X_t} = \frac{k_f X_t^2}{X_t^2 + K_d} - k_d X_t + R_{bas}+ X_t \dot{L}_t^{\alpha, \beta} , \qquad  X_0 = x,
\end{equation}
where $L_t^{\alpha, \beta}$ is an  asymmetric stable L\'{e}vy motion with the jump measure $\nu_{\alpha, \beta}$,  for the non-Gaussianity index $0<\alpha<2$ and skewness index $-1 \leq \beta \leq 1$. This L\'{e}vy motion  will be recalled in the next section. Here we assume that the generating triplet of asymmetric stable L\'{e}vy motion is $(0, 0, \varepsilon\nu_{\alpha, \beta})$, where $\varepsilon$ is the noise intensity. The noise intensity plays an important role in the noise source \cite{AAD2008,AAD2009,AAD2012,AAD2016}. We will discuss the effects of noise intensity  $\varepsilon$ in section 5.  In stochastic dynamics, it is customary to denote a state variable in a capital letter, with time dependence as subscript. The `$x$' here and hereafter denotes the initial concentration for the transcription activator factor or \textcolor{blue}{TF-A} monomer in this gene regulatory system.

Under the interaction of the potential field $U$ and  these fluctuations, the concentration of the \textcolor{blue}{TF-A}  monomer may exit from the domain $D = (0, x_u)$ ( the low concentration domain ). Our goal is to quantify the effects of asymmetric L\'evy noise on the dynamical behaviors of the \textcolor{blue}{TF-A} monomer concentration in this model. We focus on the likelihood for  the \textcolor{blue}{TF-A}  monomer concentration transitions from the low concentration domain $D$  to the high concentration domain $E=[x_u, +\infty)$ (the high concentration domain), via analyzing two deterministic quantities:  the mean residence time (also called mean first exit time) in the domain $D$ before first exit, and  the likelihood of first escape from $D$ through the right side ( i.e., becoming high concentration).  It is desirable to focus mainly on the high \textcolor{blue}{TF-A}  monomer concentration, since that corresponds to the high degree of activity. That is, high concentration indicates effective transcription and translation activities.\\


\section{Asymmetric stable L\'{e}vy motion}

The aforementioned  asymmetric stable L\'{e}vy motion   $L_t^{\alpha, \beta}$  is an appropriate model for non-Gaussian fluctuations with bursts or jumps.  The parameter $\alpha$ is the non-Gaussianity index ($0<\alpha < 2$)  and  $\beta$ is the skewness index  ($-1 \leq\beta\leq 1$).  A scalar L\'{e}vy motion has jumps that are characterized by a Borel measure $\nu$,  defined on the real line $\mathbb{R}^{1} $ and concentrated on $\mathbb{R}^{1}\backslash \{0\}$. The jump measure  $\nu$  satisfies the following condition:
\begin{equation}
  \int_{\mathbb{R}^{1}\backslash \{0\}}(|y|^{2} \wedge 1) \nu(dy) < \infty. \notag
\end{equation}

 The asymmetric stable L\'{e}vy motion   $L_t^{\alpha, \beta}$ is a stochastic process defined on a sample space $\Omega$ equipped with probability $\mathbb{P}$. It has independent and stationary increments, together with stochastically continuous sample paths:   for each  $s$, $L_t^{\alpha, \beta} \rightarrow $ $L_s^{\alpha, \beta}$ in probability. This means  for every  $\delta>0$, $\mathbb{P}$($\mid$$L_t^{\alpha, \beta}-L_s^{\alpha, \beta}$$\mid$ $>$ $\delta$)$\rightarrow 0$,  as $t\rightarrow s$.


 The jump measure,  which describes jump intensity and size for sample paths,  for the asymmetric L\'{e}vy motion  $L_t^{\alpha, \beta}$  is \cite{Applebaum2009,Duan2015},
\begin{equation} \label{eq:3}
 \nu_{\alpha, \beta}(dy)=\frac{C_1 I_ {\{0<y<\infty\}}(y)+C_2 I_{\{-\infty<y<0\}}(y)}{\mid y\mid ^{1+\alpha}}dy ,
\end{equation}
with $ C_1 =\frac{H_\alpha (1+\beta)}{2}$, $C_2= \frac{H_\alpha (1-\beta)}{2}.$
When $\alpha=1$,  $H_\alpha=\frac{2}{\pi}$; when $\alpha\neq 1$, $H_\alpha=\frac{\alpha(1-\alpha)}{\Gamma(2-\alpha)\cos(\frac{\pi \alpha}{2})}.$\\

Especially for  $\beta=0$,  this is the     symmetric stable L\'{e}vy motion,  which is usually denoted by  $L_t^\alpha  \triangleq  L_t^{\alpha, 0}$.
The well-known Brownian motion  $B_t$  may be regarded  as a special case (i.e., Gaussian case) corresponding to $\alpha=2$ (and $\beta=0$); see  \cite{Duan2015}.

\begin{figure}[!htb]
\subfigure[]{ \label{Fig.sub.21}
\includegraphics[width=0.45\textwidth]{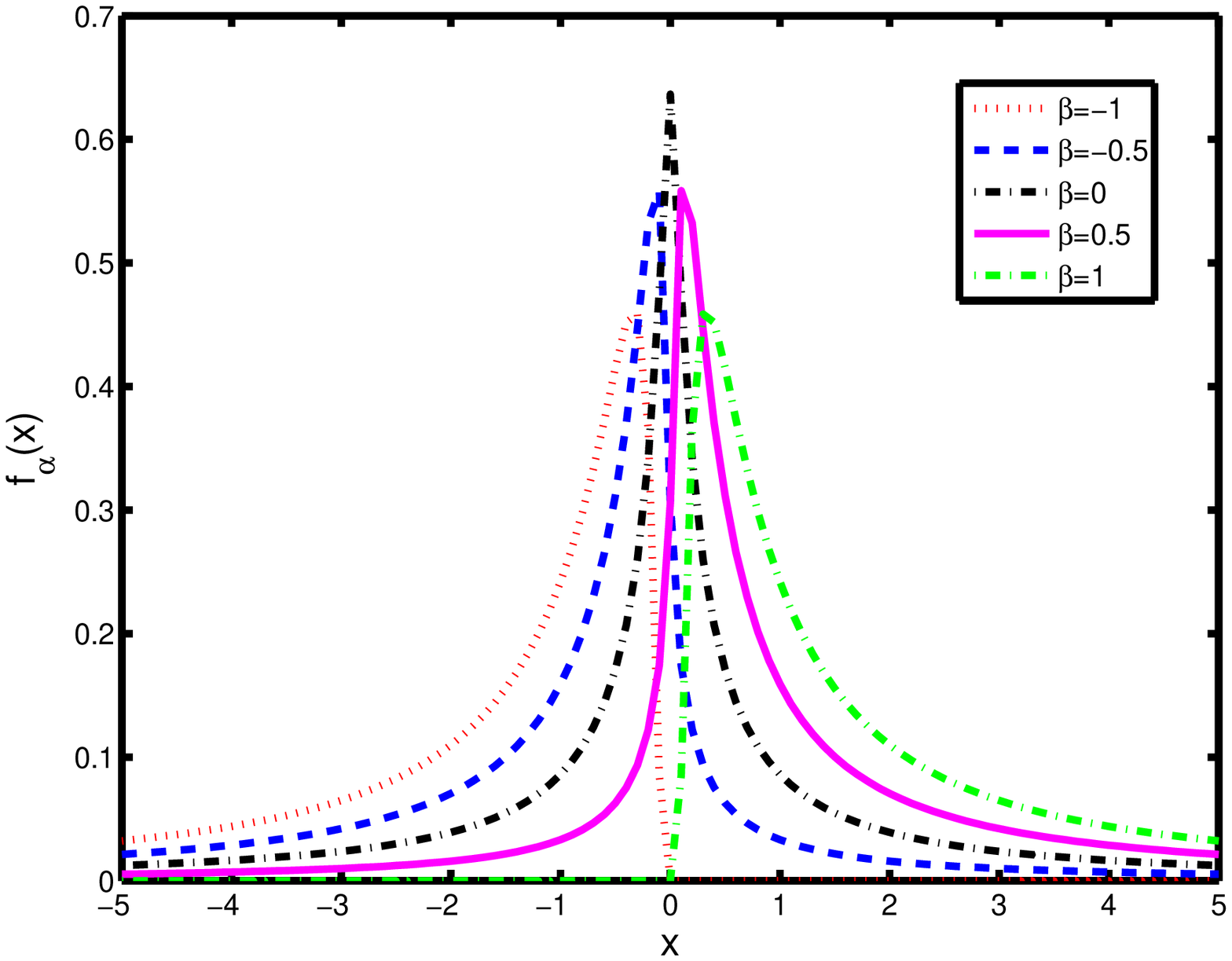}}
\subfigure[]{ \label{Fig.sub.22}
\includegraphics[width=0.45\textwidth]{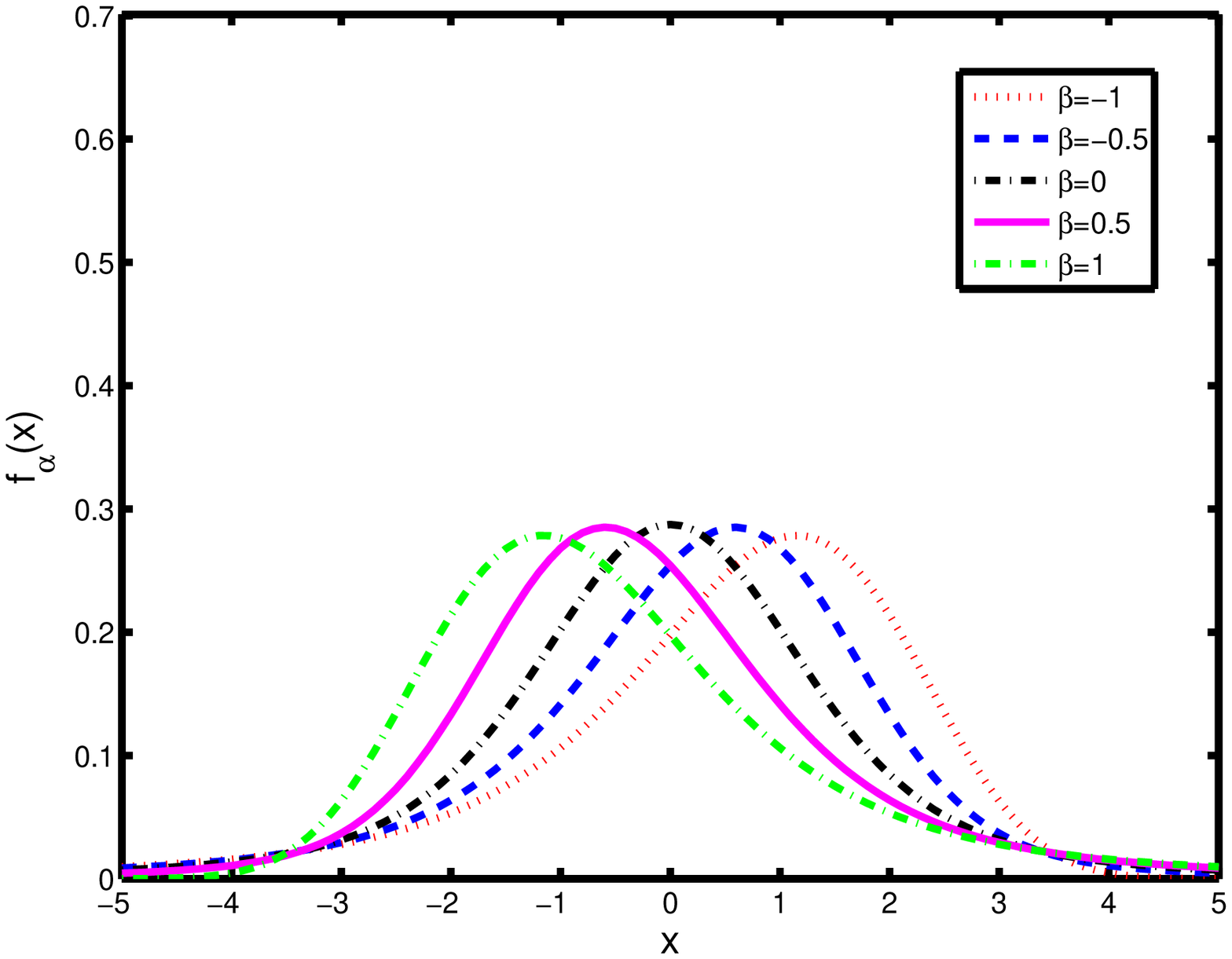}}
\caption{(Color online) Probability density functions for asymmetric stable L\'evy motion
$L_t^{\alpha, \beta}$ at $t=1$  for various skewness index $\beta$: (a) $\alpha=0.5$. (b) $\alpha=1.5$.  The asymmetry is clearly seen when $\beta \neq 0$. }
 \label{Fig.2}
\end{figure}

We can see a clear difference to the symmetric case from Figure \ref{Fig.2},  which shows the probability density functions for  $L_t^{\alpha, \beta}$ at $t=1$ for various $\alpha, \beta$.

For the stable L\'{e}vy motion with the jump measure in  \eqref{eq:3}, the number of larger jumps for  small $\alpha$ $(0< \alpha <1)$ are more than that for large  $\alpha$ $(1< \alpha <2)$, while the number of smaller jumps for $0< \alpha <1$ are less than that for $1< \alpha <2$, as known in  \cite{Sato1999}.\\

 \bigskip
To quantify the likelihood for transcription for the stochastic genetic regulatory system  \eqref{eq:2} under  \emph{asymmetric}  (i.e., non-symmetric)  stable L\'evy noise,
 we will compute   two deterministic quantities,  the mean first exit time (MFET) and the first escape probability (FEP). They are solutions of nonlocal integral-differential equations, i.e.,  \eqref{eq:4} and  \eqref{eq:6}, respectively, in   Appendix at the end of this paper.

\section{Gene regulation with synthesis rate under additive asymmetric L\'{e}vy fluctuations}
In this section, we first present the numerical schemes for solving the mean exit time $u$ and escape probability $p$,  then conduct numerical simulations to gain insights about likelihood for transcriptions modeled by \eqref{eq:2}.

\subsection{Numerical algorithms}

 For the stochastic differential equation  \eqref{eq:2} of  genetic regulation system with synthesis rate under   asymmetric L\'{e}vy noise,  we present a numerical scheme to solve the following deterministic nonlocal integral-differential equation,  \eqref{eq:4}  in   Appendix,   in order to   get the mean first exit time $u$.
 \begin{eqnarray}
  Au(x) &=& -1, \qquad  x \in D,\notag\\
  u(x)  &=& 0, \qquad  x \in D^{c}.
\end{eqnarray}
Here $ D^{c}$ is the complement set of $D$ in $\mathbb{R}^{1}$.

 The generator $A$ for  the stochastic differential equation \eqref{eq:2} with asymmetric  stable L\'{e}vy motion   is   \cite{Duan2015,Sato1999}
\begin{equation}
Au(x)=(f(x)+ \varepsilon M_{\alpha, \beta})u'(x)  + \varepsilon  \int_{\mathbb{R}^{1}\backslash \{0\}}[u(x + y) - u(x) - I_{\{|y|<1\}}(y)yu'(x)]\nu_{\alpha, \beta}(dy),  \label{eq:7}
\end{equation}
with $\nu_{\alpha, \beta}(dy)=\frac{C_1 I_ {\{0<y<\infty\}}(y)+C_2 I_{\{-\infty<y<0\}}(y)}{\mid y\mid ^{1+\alpha}}dy,$
$ C_1 =\frac{H_\alpha (1+\beta)}{2}$ and  $C_2= \frac{H_\alpha (1-\beta)}{2}.$
When $\alpha=1$, $H_\alpha=\frac{2}{\pi}$; when $\alpha\neq 1$, $H_\alpha=\frac{\alpha(1-\alpha)}{\Gamma(2-\alpha)\cos(\frac{\pi \alpha}{2})}.$
Additionally,
$$
M_{\alpha, \beta} =
\left \{
  \begin{array}{ll}
    \frac{C_1-C_2 }{1-\alpha}, & \hbox{$\alpha \neq 1$,} \\
    \ (\int_{1}^{\infty}{\frac{\sin(x)}{x^2}}dx+\int_{0}^{1}{\frac{\sin(x)-x}{x^2}}dx )(C_2-C_1), & \hbox{$\alpha = 1$.}
  \end{array}
\right.
$$
The MFET $u$ satisfies the following equation:
\begin{eqnarray}
& & (f(x)+ \varepsilon M_{\alpha, \beta})u'(x)  \nonumber \\
&+&  \varepsilon \int_{\mathbb{R}^{1}\backslash \{0\}}[u(x+y) - u(x) - I_{\{|y|<1\}}(y)yu'(x)]\frac{[C_1 I_{\{0<y<\infty\}}(y)+C_2 I_{\{-\infty<y<0\}}(y)]}{\mid y\mid^{1+\alpha}}dy \nonumber  \\
&=& -1.     \label{eq:8}
\end{eqnarray}
On an open interval $D=(a,b)$, we make a coordinate  conversion $x=\frac{b-a}{2}s + \frac{b+a}{2}$ for $s \in [-1, 1]$ and $y=\frac{b-a}{2}r$,  to get finite difference discretization for $Au(x)=-1$ as in \cite{Gao2014}:          \\
\begin{equation}  \label{eq:9}
\begin{aligned}
(\frac{2}{b-a})f(\frac{b-a}{2}s + \frac{b+a}{2}+ \varepsilon M_{\alpha, \beta})u'(s) + \varepsilon (\frac{2}{b-a})^{\alpha} \int_{\mathbb{R}^{1}\backslash \{0\}}[u(s+ r) - u(s) - I_{\{|r|<1\}}(r)ru'(s)]              \\
\frac{[C_1 I_{\{0<r<\infty\}}(r)+C_2 I_{\{-\infty<r<0\}}(r)]}{\mid r \mid ^{1+\alpha}}dr=-1.
\end{aligned}
\end{equation}
With the numerical simulation via  \eqref{eq:9}, we obtain the MFET $u$ for the stochastic gene regulation model  \eqref{eq:2}.

A similar scheme is used  for numerical simulation for \eqref{eq:6}  in   Appendix, to get the first escape probability $p$.

\subsection{Numerical experiments}

\medskip
We summarize major numerical simulation results below, and indicate their relevance to the likelihood for gene transcriptions. We highlight the peculiar dynamical differences with the case of symmetric stable L\'evy noise ($\beta =0$) in \cite{Zheng2016}.

As we take domain $D=(0, x_u)$ to be in the low concentration region, a smaller MFET indicates higher likelihood for gene transcription (and vice versa), and a larger FEP means higher likelihood for gene transcription (and vice versa). Both MFET $u$ and FEP $p$ reflect  the interactions   between nonlinear vector field $f$ and the asymmetric stable L\'evy noise $L_t^{\alpha, \beta}$. \\

\begin{figure}[!htb]
\subfigure[]{ \label{Fig.sub.31}
\includegraphics[width=0.45\textwidth]{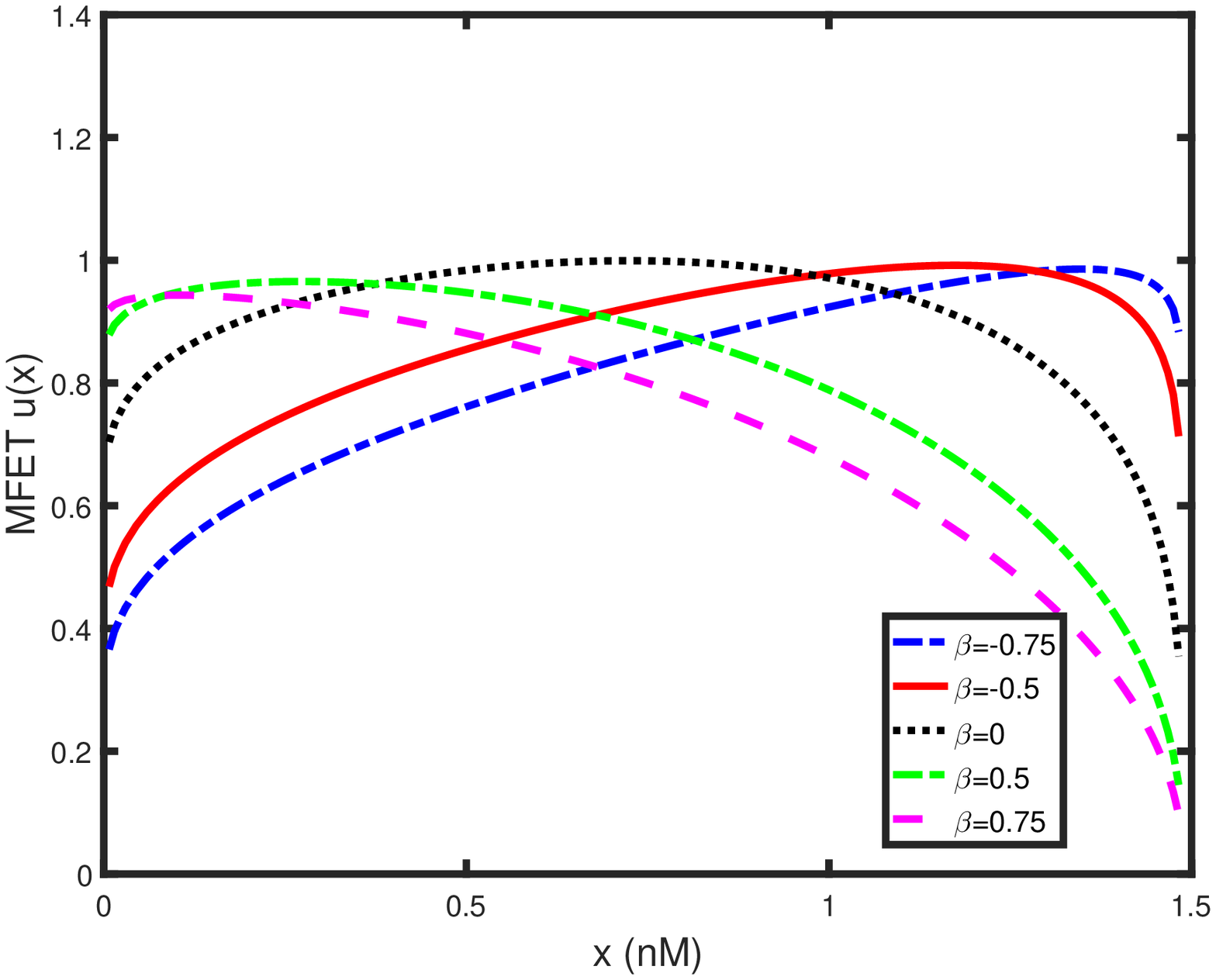}}
\subfigure[]{ \label{Fig.sub.32}
\includegraphics[width=0.45\textwidth]{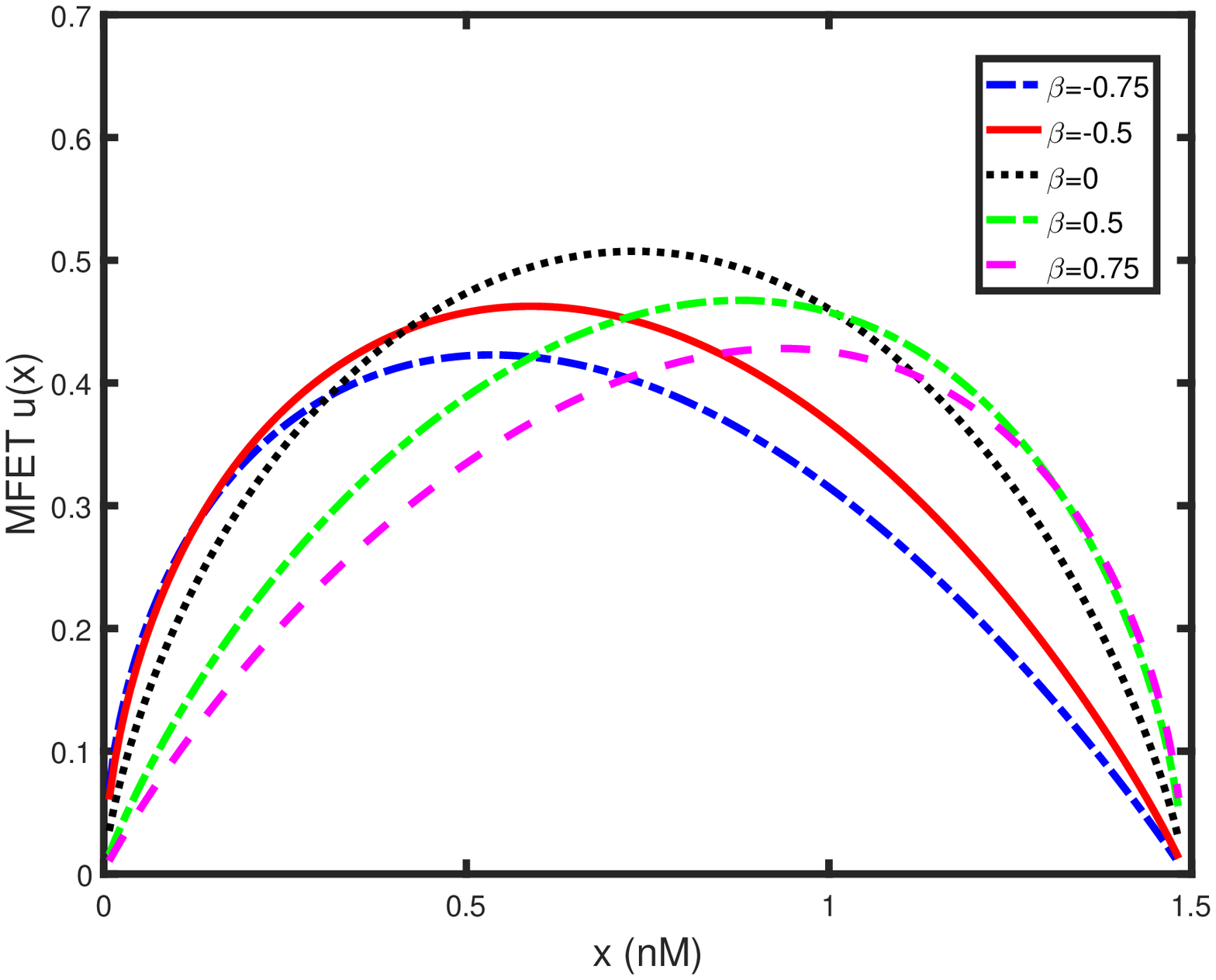}}

\caption{ (Color online) Mean first exit time (MFET)  $u(x)$ as a function of initial concentration $x$ in the low concentration domain $D = (0, 1.48971)$. Effect of skewness index $\beta$ on the MFET: (a)  $\alpha=0.5$. (b)  $\alpha=1.5$.}
 \label{Fig_3}
\end{figure}

\begin{figure}[!htb]
\subfigure[]{ \label{Fig.sub.41}
\includegraphics[width=0.45\textwidth]{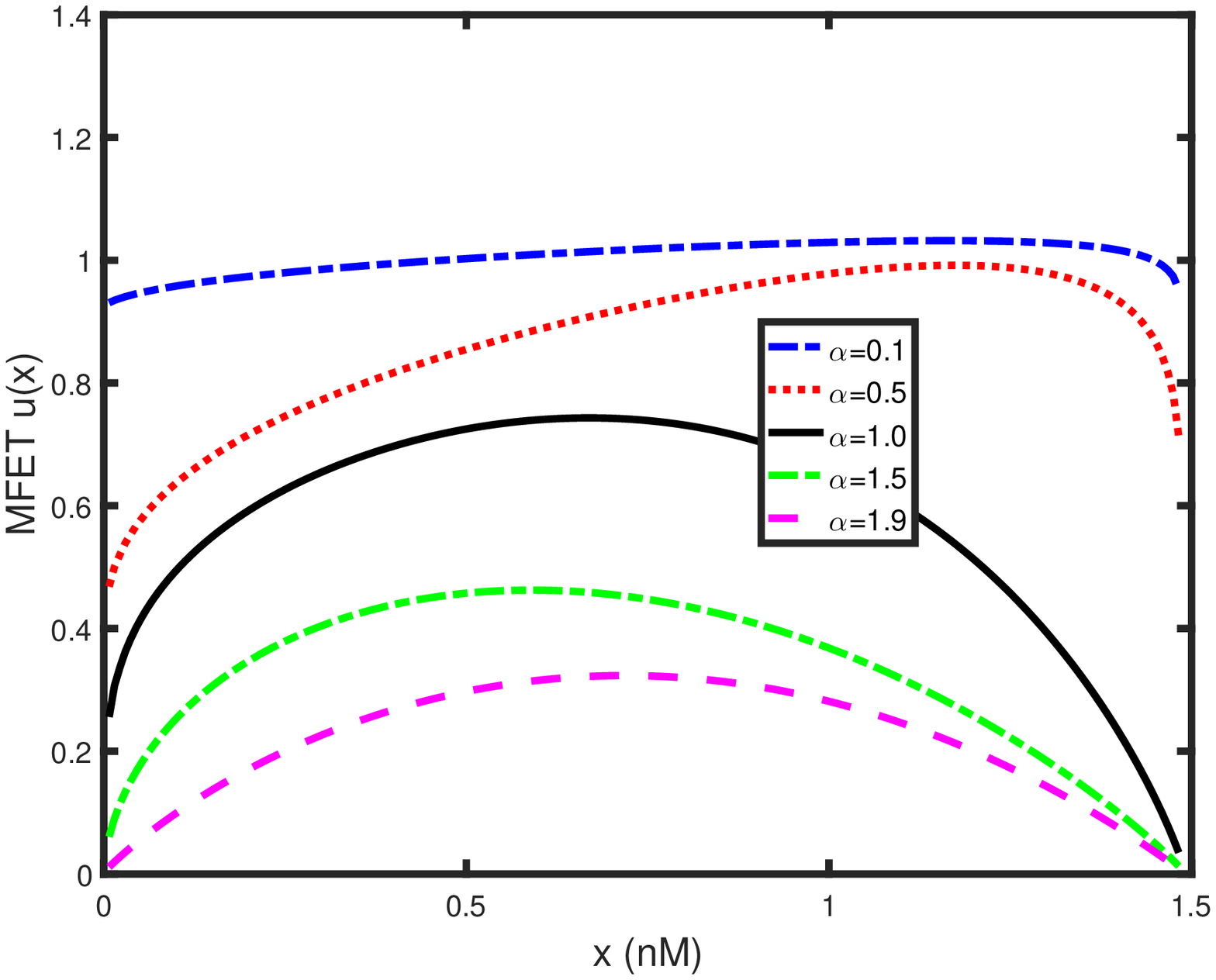}}
\subfigure[]{ \label{Fig.sub.42}
\includegraphics[width=0.45\textwidth]{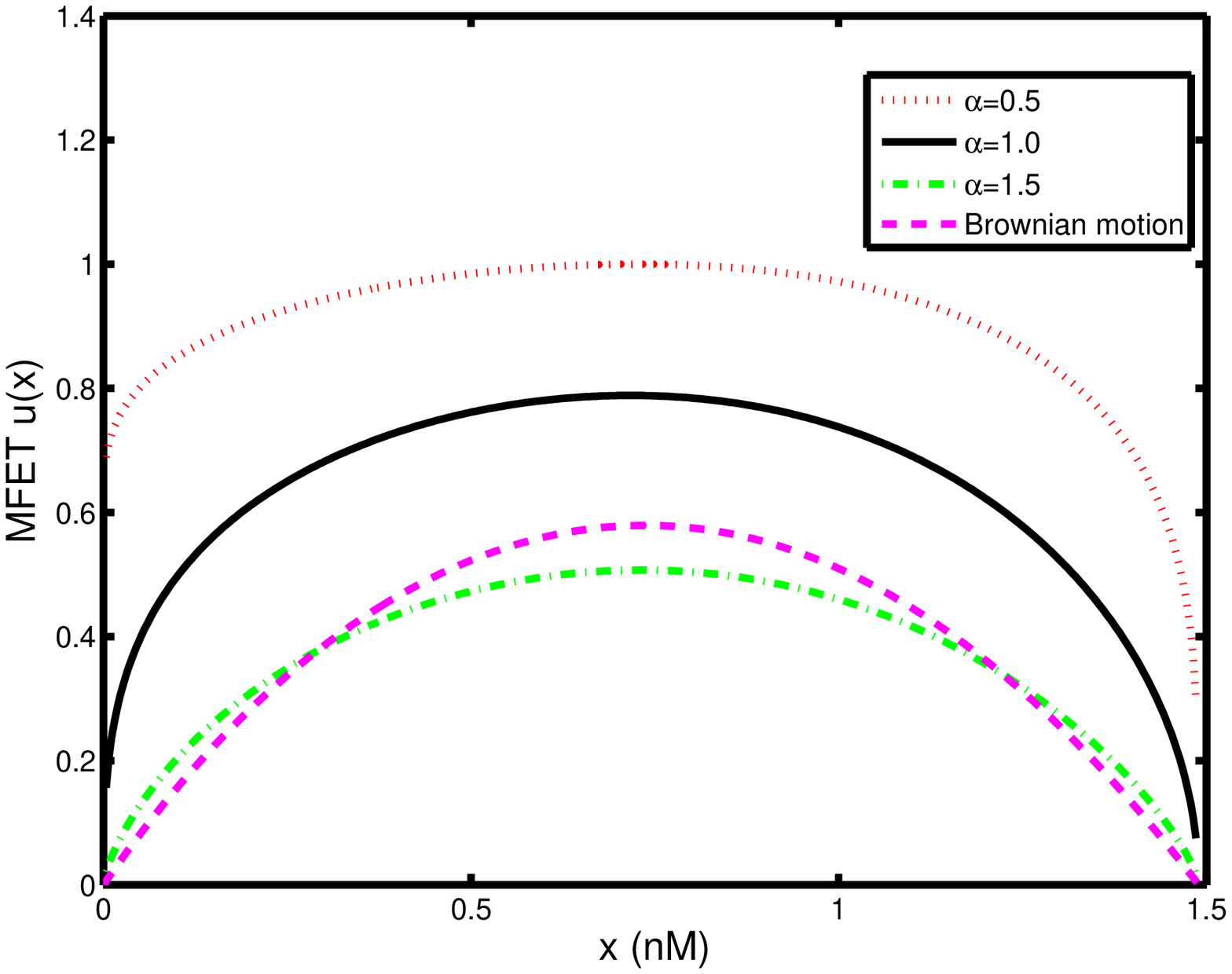}}
\caption{ (Color online) Mean first exit time (MFET)  $u(x)$ as a function of initial concentration $x$ in the low concentration domain $D = (0, 1.48971)$.
Effect of non-Gaussianity index  $\alpha$ on the MFET:  (a)  $\beta=-0.5$. (b)  $\beta = 0$. }
\label{Fig_4}
\end{figure}

\begin{figure}[!htb]
\subfigure[]{ \label{Fig.sub.51}
\includegraphics[width=0.45\textwidth]{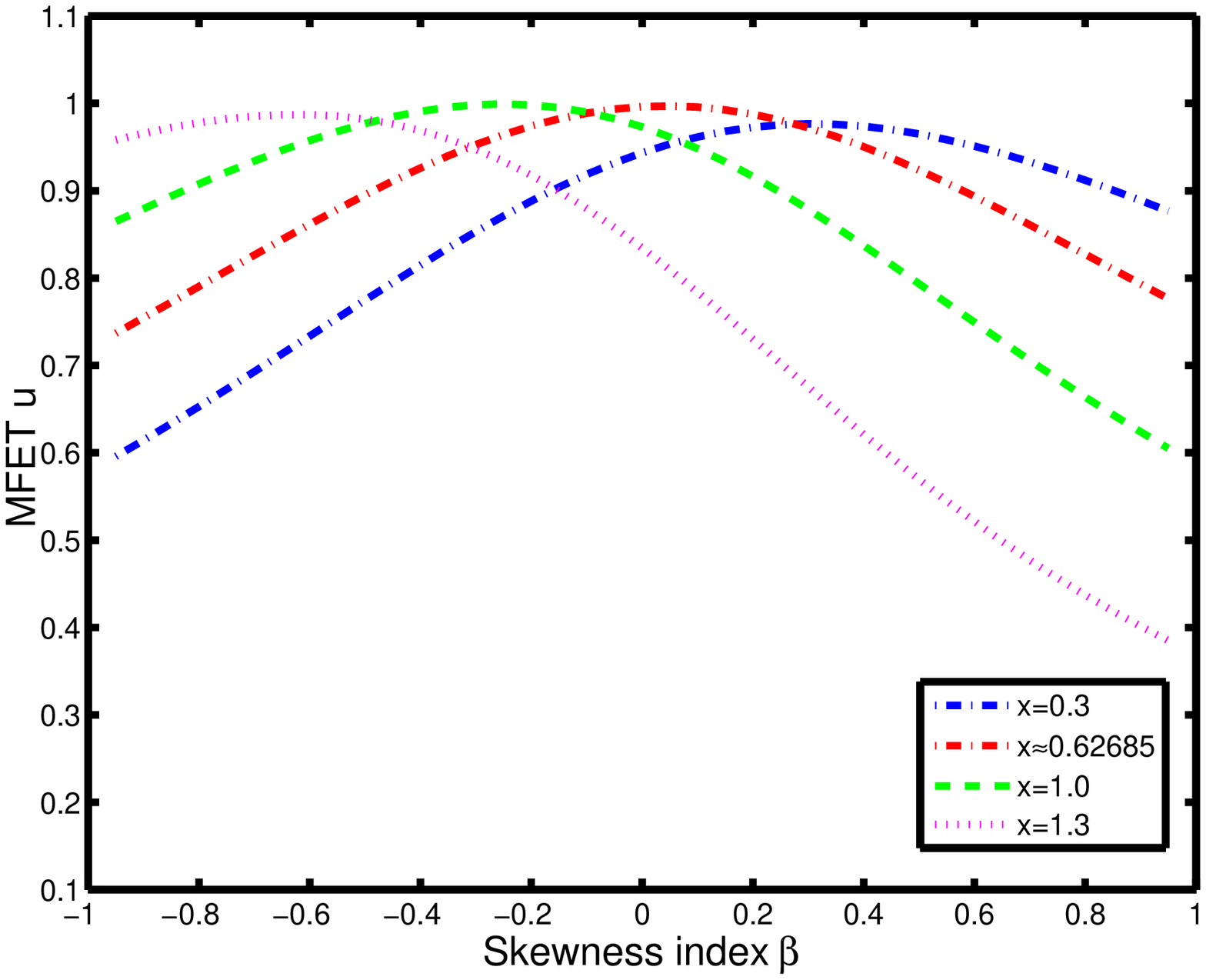}}
\subfigure[]{ \label{Fig.sub.52}
\includegraphics[width=0.45\textwidth]{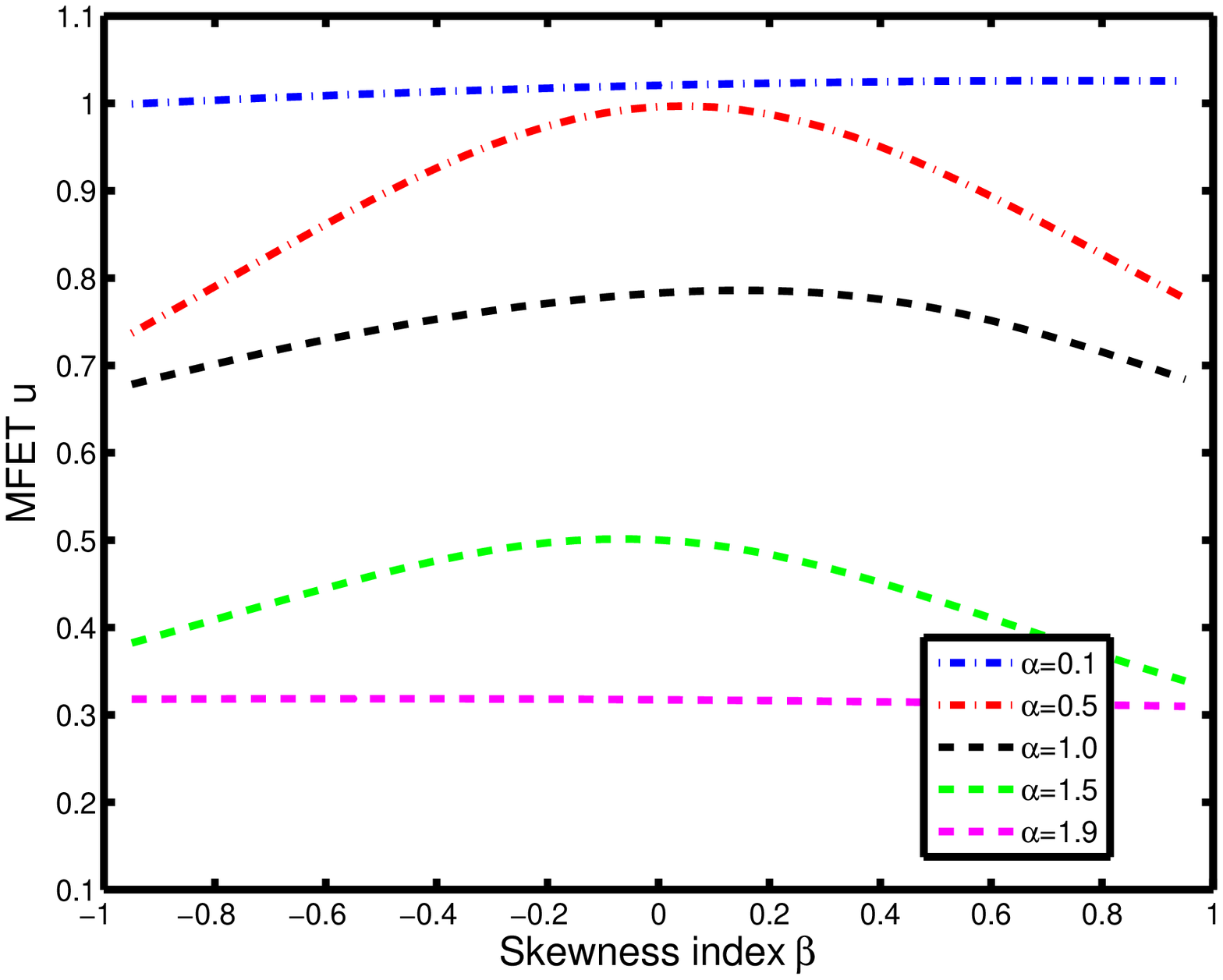}}
\caption{ (Color online) Mean first exit time (MFET) $u $ as a function of  skewness index $\beta$.  (a) Effect of initial concentrations  $x$  and $\beta$ on the MFET:  $\alpha=0.5$.  (b) Effect of   $\alpha$ and $\beta$ on the MFET at the lower stable concentration state:  $x =x_{-}\approx 0.62685$.}
 \label{Fig_5}
\end{figure}

Figure \ref{Fig_3} shows the impact of the skewness  index $\beta$   on  MFET, for $\alpha=0.5$ and $\alpha=1.5$ . When $-1<\beta<0$, MFET increases firstly then decreases, but for $0<\beta<1$, MFET decreases in the whole interval. This indicates that the asymmetry of the noise (characterized by $\beta$) plays an important role in the dynamical system: Increasing positive asymmetry leads to higher likelihood for gene transcription, while for negative asymmetry there is a minimum likelihood for transcription ($\alpha=0.5$). But for  $\alpha=1.5$, MFET increases to the maximum and then decreases to $0$, i.e., there is a minimum likelihood for transcription for all asymmetry index $\beta$. Meanwhile, we observe that for $\beta<0$, MFET decreases earlier than that for $\beta>0$. We also observe a peculiar feature.  With   $\alpha <1$,   the MFET reaches the  maxima value  (i.e., the least likelihood for transcription) near the exit boundary $x_u=1.48971$   for negative $\beta$; while with   $\alpha >1$,  the MFET reaches the  maxima value near (i.e., the least likelihood for transcription) the exit boundary $x_u=1.48971$   for positive $\beta$. This indicates that the skewness index $\beta$ may function as a tuning parameter for transcription.

 Figure \ref{Fig_4} shows that  when $\beta$ is fixed, the MFET values decrease with the increasing $\alpha$, i.e., the likelihood for gene transcription increases with increasing $\alpha$.
 In comparison, Figure 4(b) contains the case with Brownian noise (i.e., corresponding to $\alpha=2, \beta=0$) and the MFET values break this monotonicity and stay roughly between those for $\alpha=1.5$ and $\alpha=1.9$.
Figure \ref{Fig_3} and Figure\ref{Fig_4} indicate that if we start  in the  low concentration, then  increasing $\alpha$ and $\beta$ values leads to the higher concentrations, corresponding to higher likelihood for transcription.

Figure \ref{Fig_5} plots the dependency of   MFET in the low concentration  on the asymmetry index $\beta$. Since the transcription behavior is particularly sensitive to initial conditions \cite{Smolen1998}, we investigate the noise effect on different initial concentrations. In the case of $\alpha=0.5$, MFET increases at first and then decreases. Different initial concentrations $x$ correspond to different maximum MFET values: By tuning the asymmetry index $\beta$ (depending on initial concentration), we can find the least likelihood for transcription. If we fix $x=0.62685$ (low stable  concentration), MFET increases and then decreases, especially  for  $\alpha =0.5 $ or $ 1.5$: By increasing non-Gaussian index $\alpha$, we can achieve higher likelihood for transcription.

\begin{figure}[!htb]
\subfigure[]{ \label{Fig.sub.61}
\includegraphics[width=0.45\textwidth]{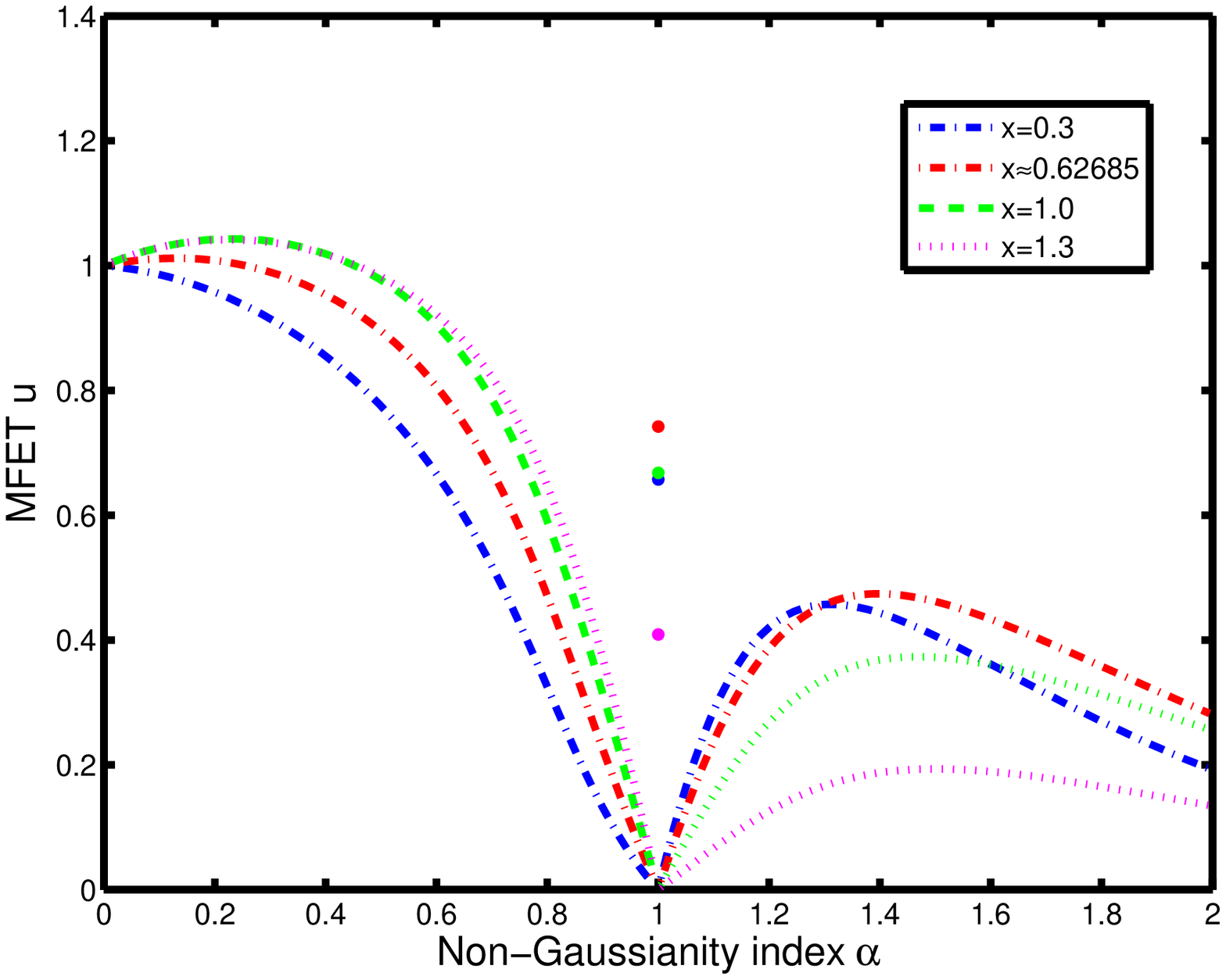}}
\subfigure[]{ \label{Fig.sub.62}
\includegraphics[width=0.45\textwidth]{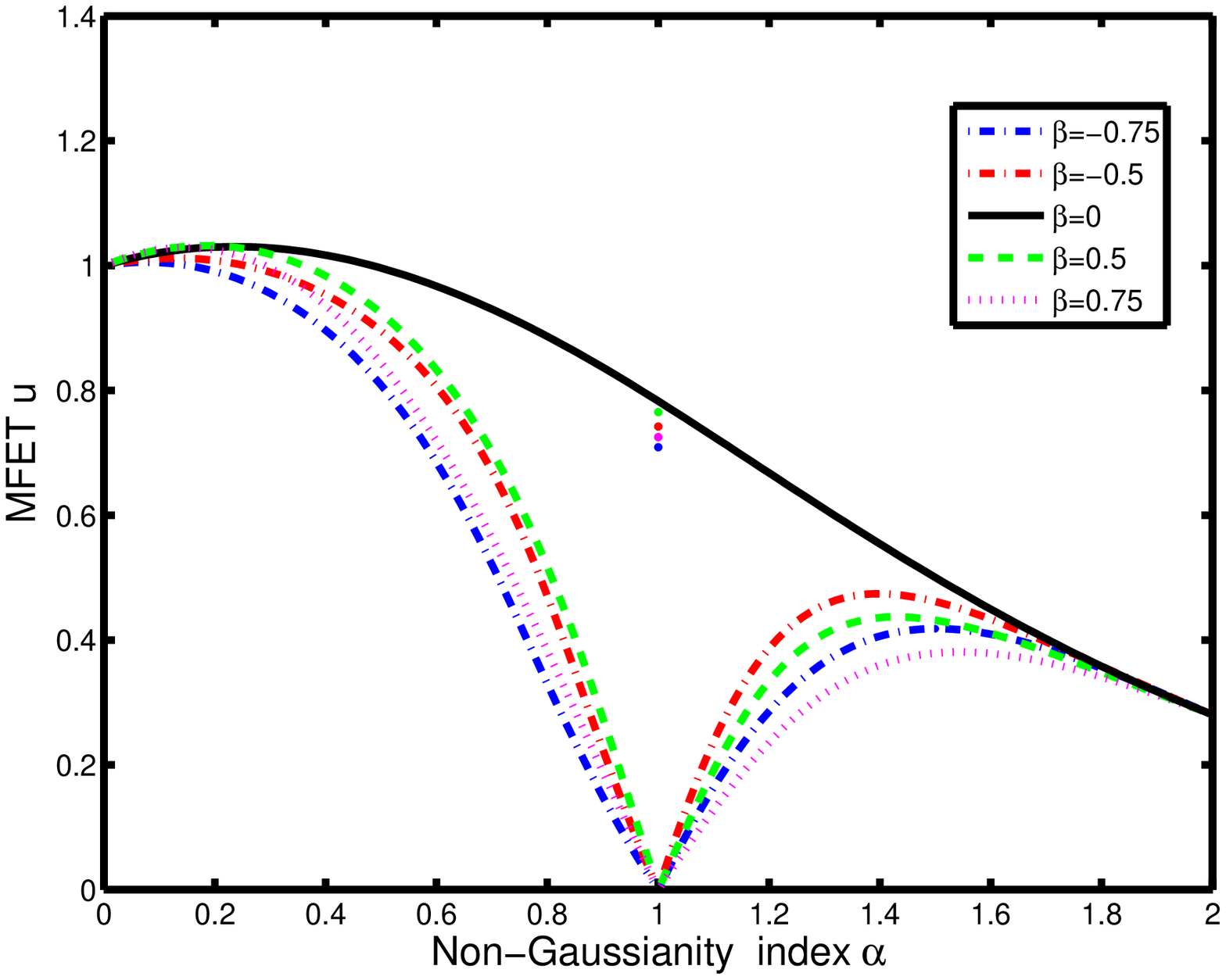}}
\caption{ (Color online) MFET $u$ as a function of $\alpha$. (a) Effect of different initial concentrations  $x$ and $\alpha$ on the MFET:  $\beta=-0.5$. (b) Effect of $\alpha$ and $\beta$ on the MFET at the lower stable concentration state: $x =x_{-}\approx 0.62685$.}
 \label{Fig_6}
\end{figure}

\begin{bfseries}
When skewness $\beta  \neq  0$: It makes a great difference on MFET for $\alpha<1$ and $\alpha>1$.
\end{bfseries}
Figure \ref{Fig_6} exhibits that, when $\beta\neq0$, MFET has a bifurcation  or  discontinuity point   at $\alpha=1$ when $\beta\neq0$.
We can see that the  MFET has a `phase  transition' or bifurcation at the critical non-Gasussian index value $\alpha=1$. This result is consistent with a theoretical analysis in \cite{Imkeller2009}. When the asymmetry index $\beta\neq0$, in the low concentration region, MFET decreases with the increasing $\alpha$ for $0<\alpha<1$, while for $1<\alpha<2$, MFET increases firstly but then decreases with the increasing $\alpha$.  In the symmetric L\'evy nose case ($\beta=0$), MFET is decreasing for all $\alpha$ (no bifurcation). Hence in the asymmetric L\'evy noise case ($\beta\neq0$): We gain higher likelihood for transcription by increasing non-Gaussian index $\alpha \in (0, 1)$, while for $\alpha \in (1, 2)$ there is a specific $\alpha_s$  leading to the minimum likelihood for transcription.

We thus observe that smaller MFET for larger non-Gaussianity index $\alpha$ and larger skewness index $\beta$. We can always achieve the minimum MFET by tuning non-Gaussianity index $\alpha$ and skewness index $\beta$. The smaller MFET means a high level of \textcolor{blue}{TF-A} ,   corresponding to a higher likelihood for gene transcription.

\begin{figure}[!htb]
\subfigure[]{ \label{Fig.sub.71}
\includegraphics[width=0.45\textwidth]{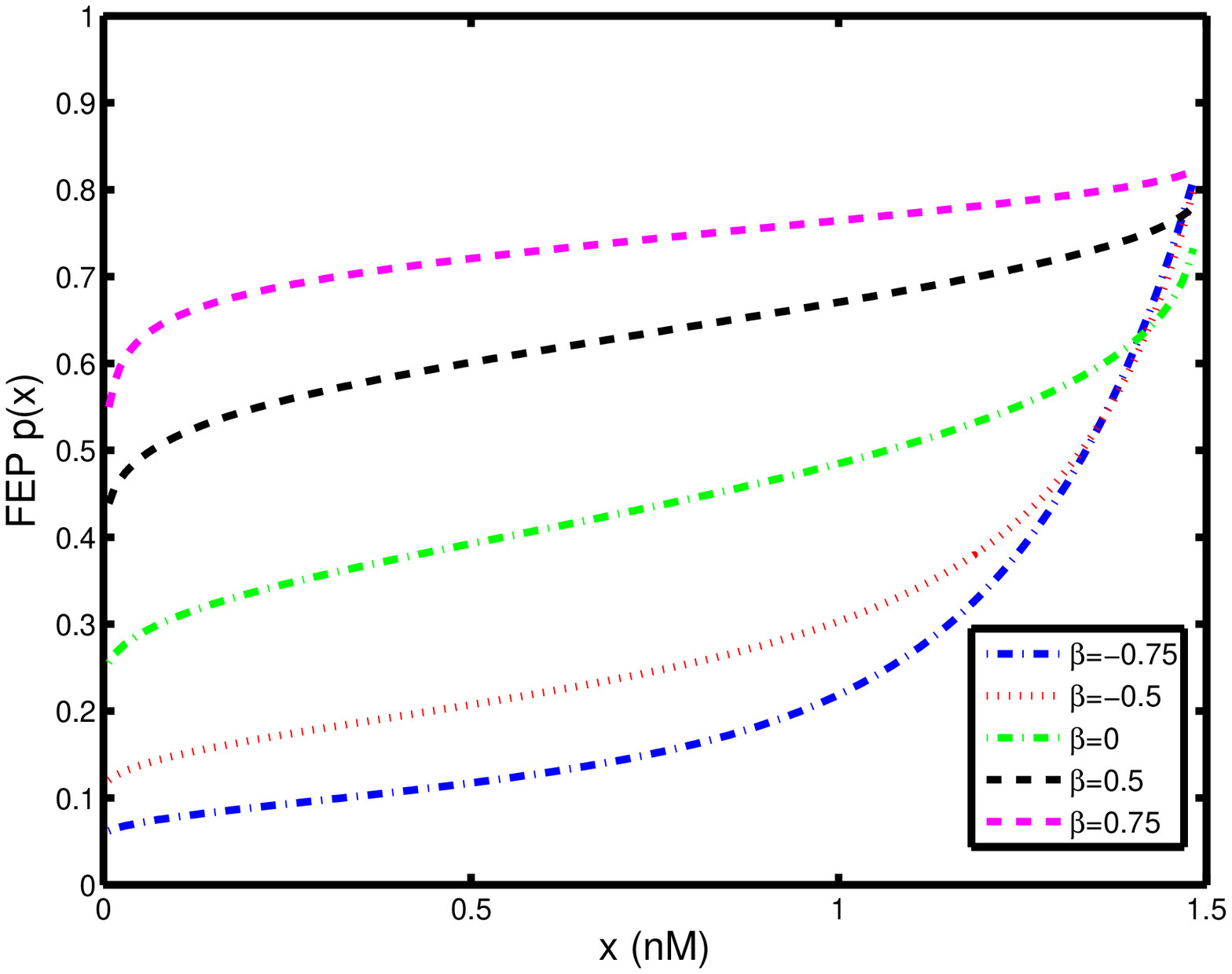}}
\subfigure[]{ \label{Fig.sub.72}
\includegraphics[width=0.45\textwidth]{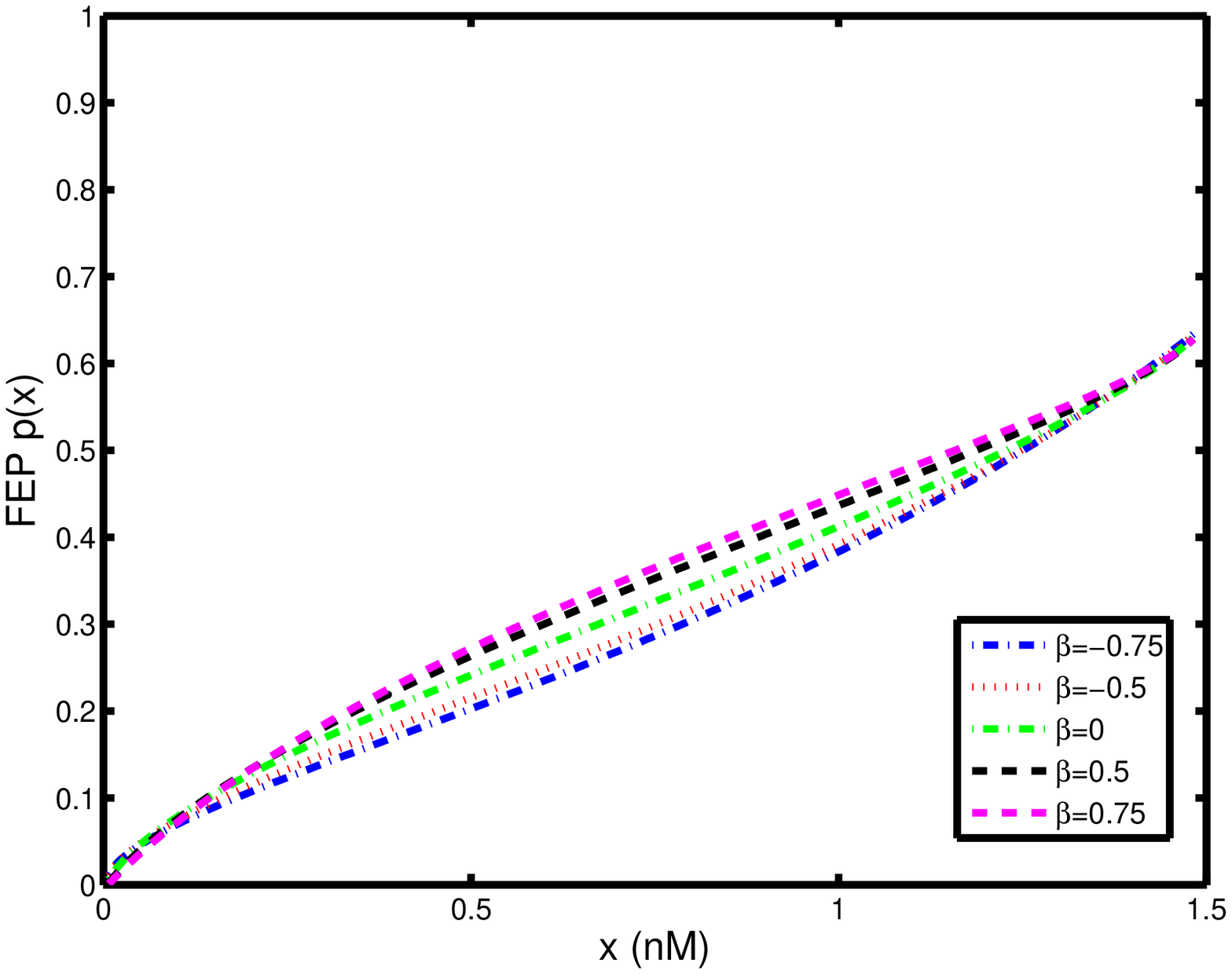}}
\caption{(Color online) FEP $p(x)$ as a function of initial concentration $x$, from $D = ( 0, 1.48971 )$ to $E=[1.48971,\infty)$.  Effect of skewness index $\beta$ on the FEP : (a) $\alpha=0.5$.  (b) $\alpha=1.5$.}
 \label{Fig_7}
\end{figure}

\begin{figure}[!htb]
\subfigure[]{ \label{Fig.sub.81}
\includegraphics[width=0.45\textwidth]{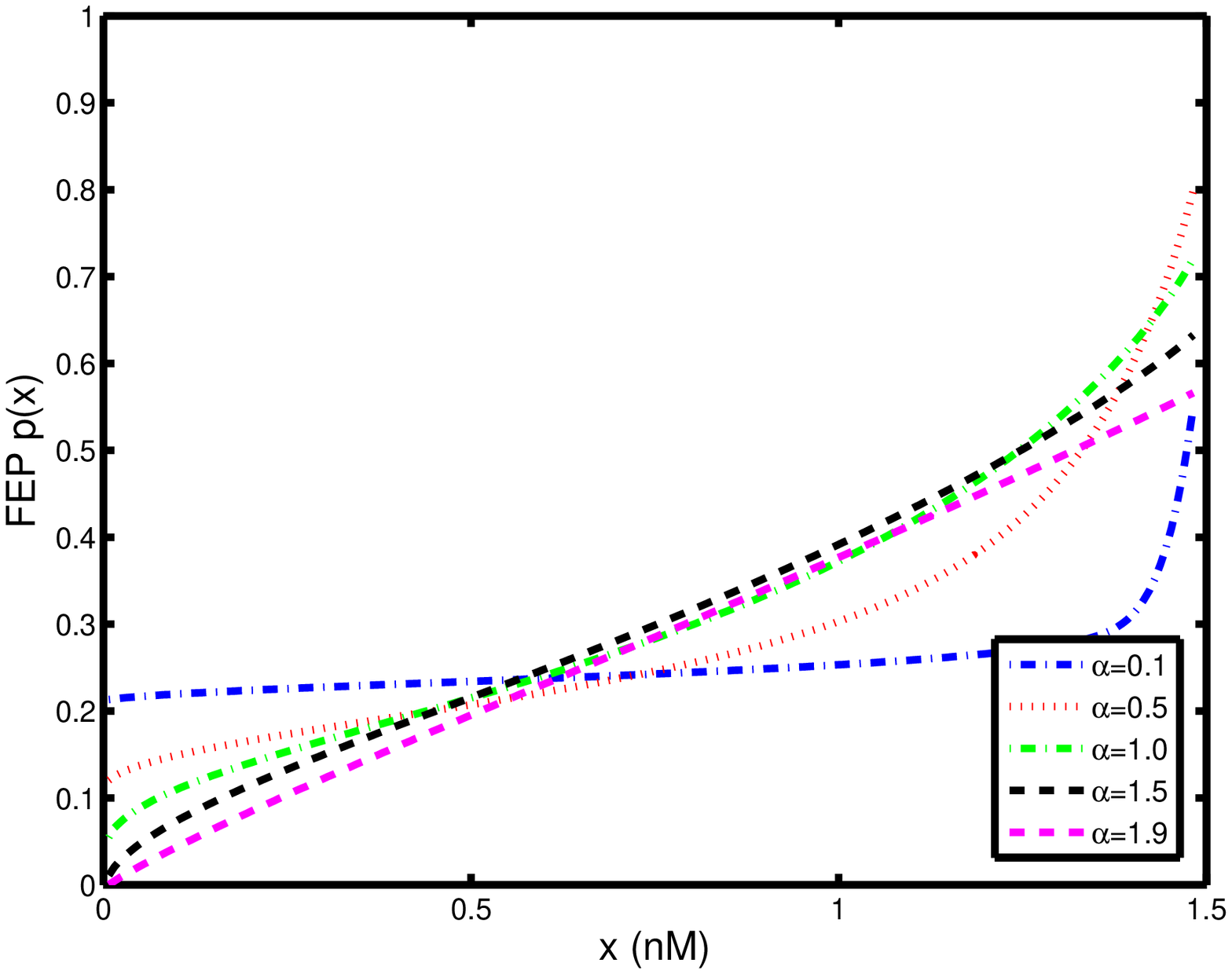}}
\subfigure[]{ \label{Fig.sub.82}
\includegraphics[width=0.45\textwidth]{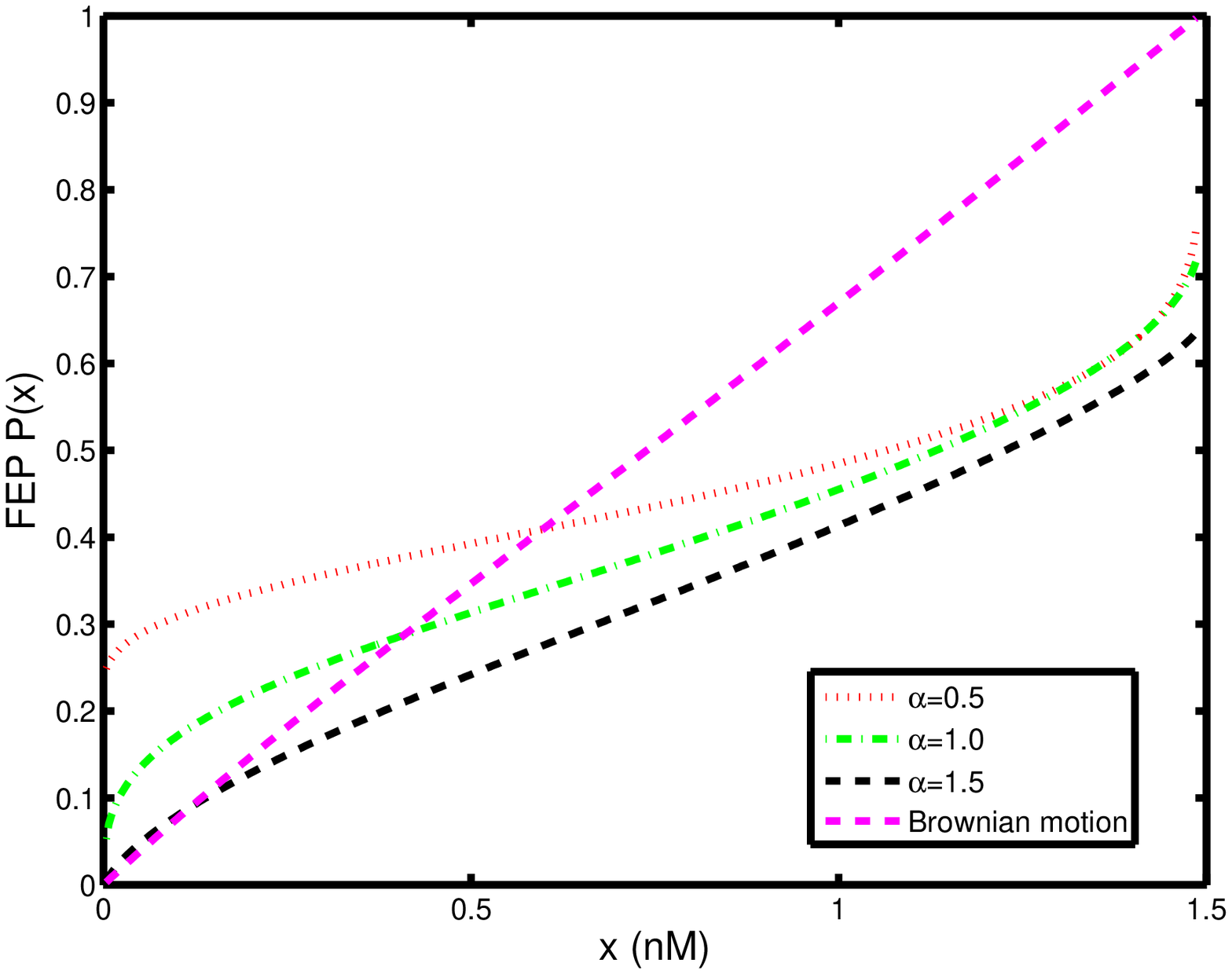}}
\subfigure[]{ \label{Fig.sub.83}
\includegraphics[width=0.45\textwidth]{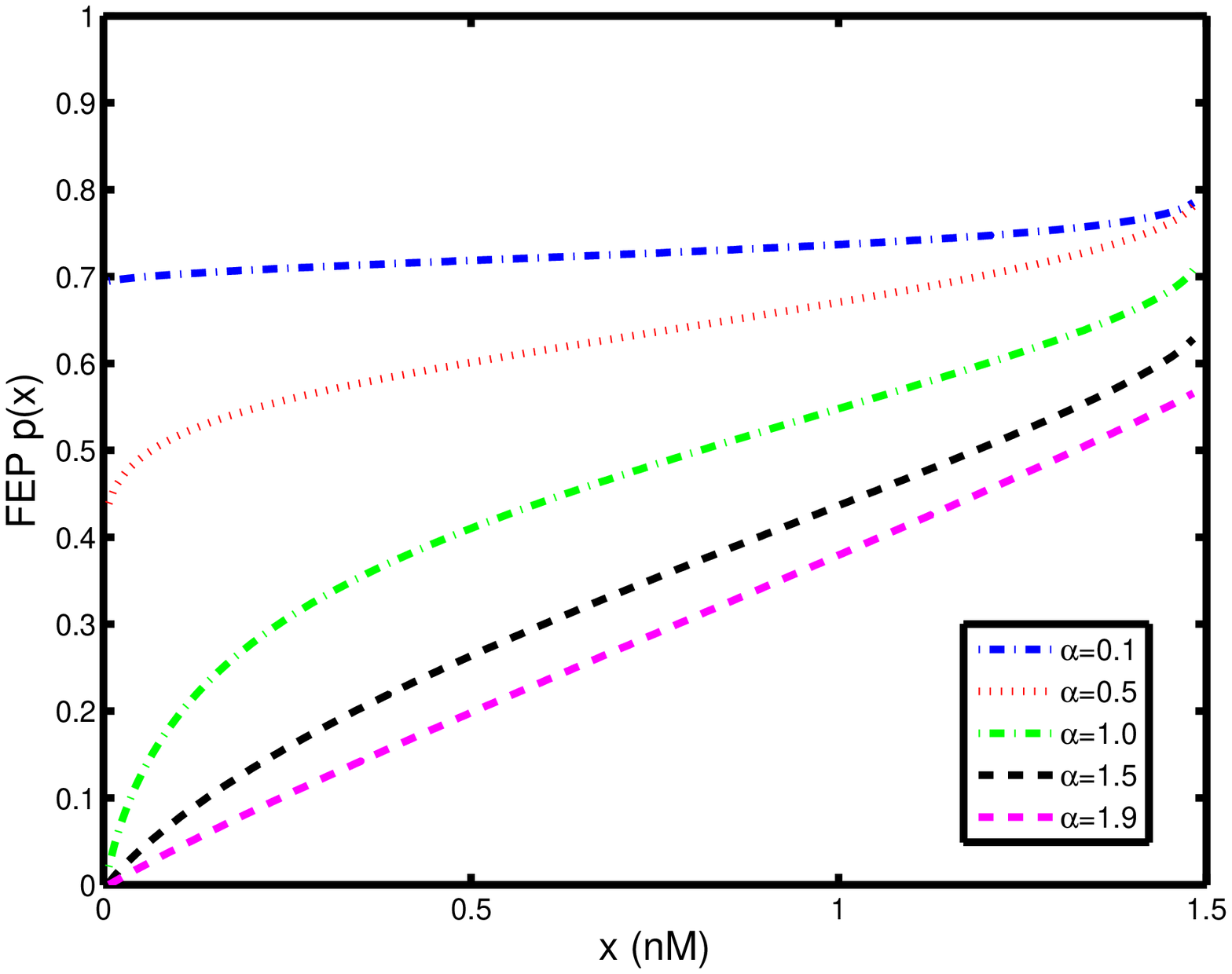}}
\caption{(Color online) FEP $p(x)$ as a function of initial concentration $x$, from $D = ( 0, 1.48971)$ to $E=[1.48971,\infty)$. Effect of non-Gaussianity index  $\alpha$ on the FEP:  (a) $\beta=-0.5$. (b) $\beta = 0$. (c) $\beta = 0.5$.}
 \label{Fig_8}
\end{figure}

\begin{figure}[!htb]
\subfigure[]{ \label{Fig.sub.91}
\includegraphics[width=0.45\textwidth]{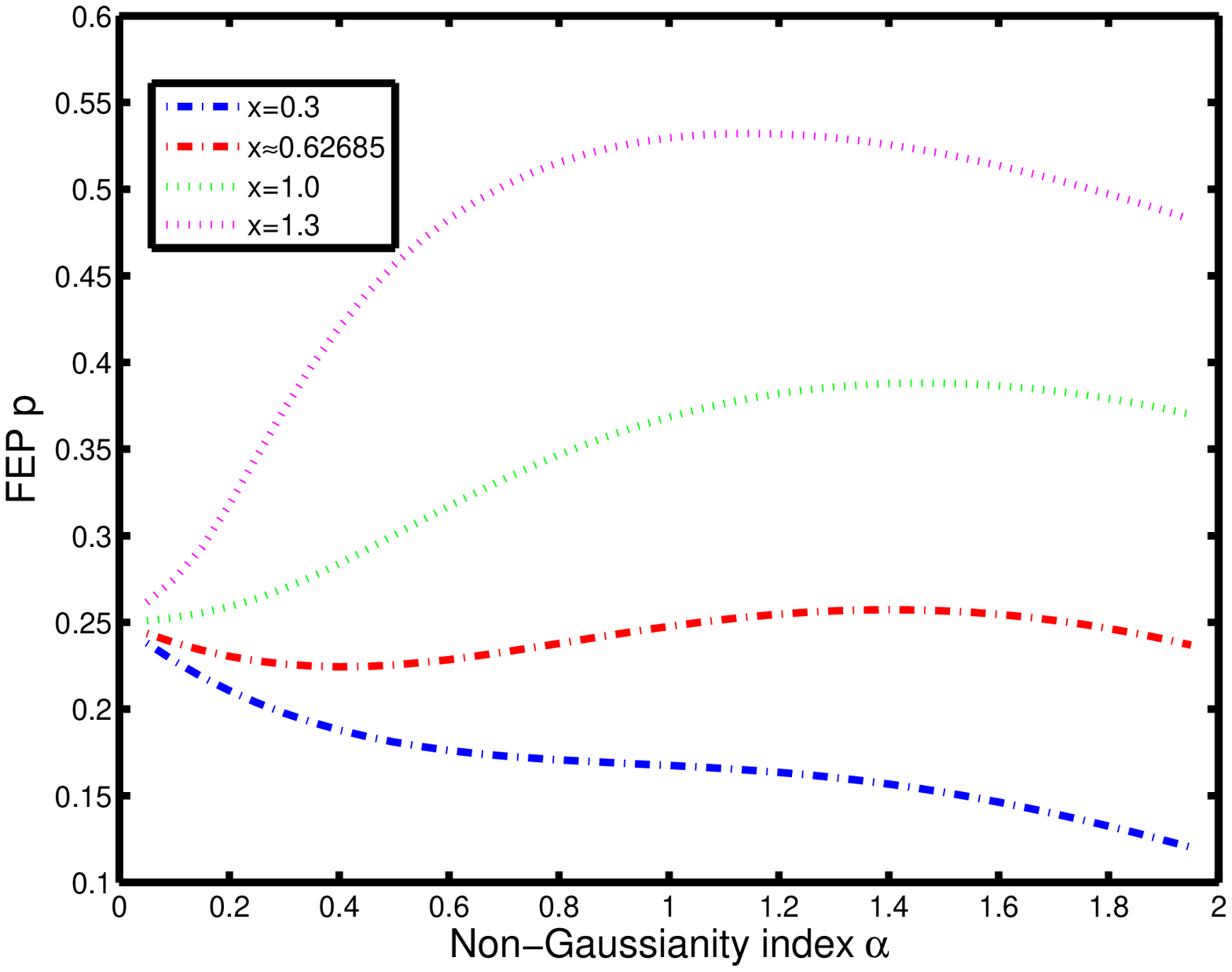}}
\subfigure[]{ \label{Fig.sub.92}
\includegraphics[width=0.45\textwidth]{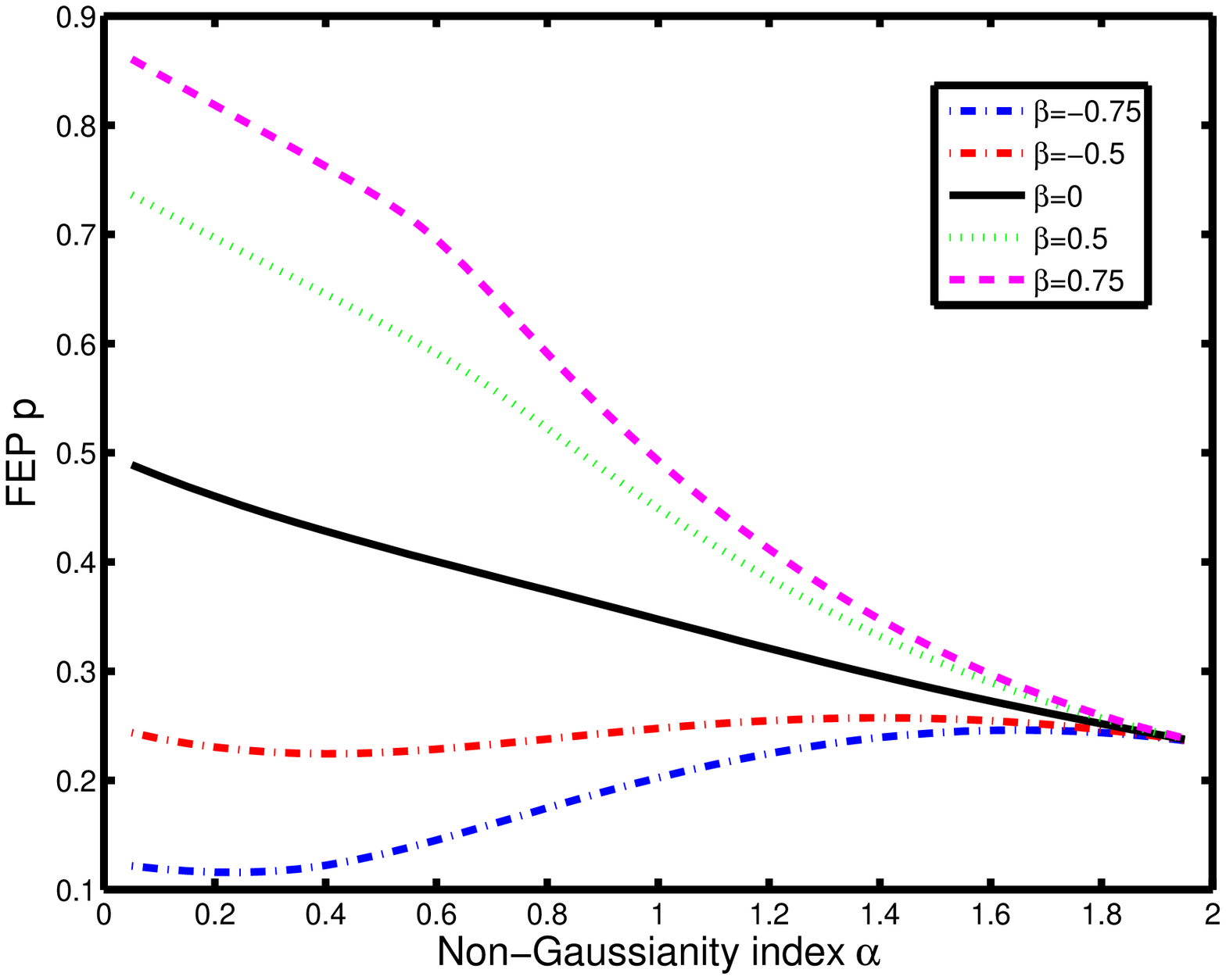}}
\caption{(Color online) FEP $p$ as a function of $\alpha$, from $D = (0, 1.48971)$ to $[1.48971,\infty)$. (a) Effect of $\alpha$ and different initial concentrations  $x$ on the FEP with  $\beta=-0.5$.  (b) Effect of $\alpha$ and $\beta$ on the FEP at the lower stable concentration state  $x =x_{-} \approx0.62685$.}
 \label{Fig_9}
\end{figure}

Figure \ref{Fig_7} demonstrates that FEP increases with the increasing $\beta$, and FEP for positive $\beta$ is larger than that for negative $\beta$.   Comparing (a) with (b), we find that FEP for  $\alpha=1.5$ increases more rapidly than that for $\alpha=0.5$.

From Figure \ref{Fig_8}, we observe that when $\beta=-0.5$, FEP corresponding to different $\alpha$ has intersection or crossover points. Before and after the intersection point, there exists an opposite relationship. When $\beta=0.5$, FEP decreases with the increasing $\alpha$. So in order to get a high likelihood of gene transcription, we can tune asymmetric index $\beta$ larger  and $\alpha$ smaller.
In comparison, for the Brownian noise case in Figure \ref{Fig_8}(b), the FEP is approximately linearly increasing in the initial concentration $x$.

As shown in Figure \ref{Fig_9},  we find that, when $\beta<0$, FEP deceases with the increasing $\alpha$ for initial concentration  $x<x_-$, then increases with the increasing $\alpha$ for  $x_-<x<x_u$.    This leads to the conclusion  that larger initial concentrations are more likely leading to the transcription. If we  consider FEP at the low concentration $x=0.62685$, we see  that when $\beta<0$, FEP increases with the increasing $\alpha$, while when $\beta\geq0$, FEP decreases with the increasing $\alpha$.  A small $\alpha$ (and $\beta>0$) or a large $\alpha$ (and $\beta<0$) contributes to large FEP (i.e., more likely for transcription).

\begin{figure}[!htb]
\subfigure[]{ \label{Fig.sub.01}
\includegraphics[width=0.45\textwidth]{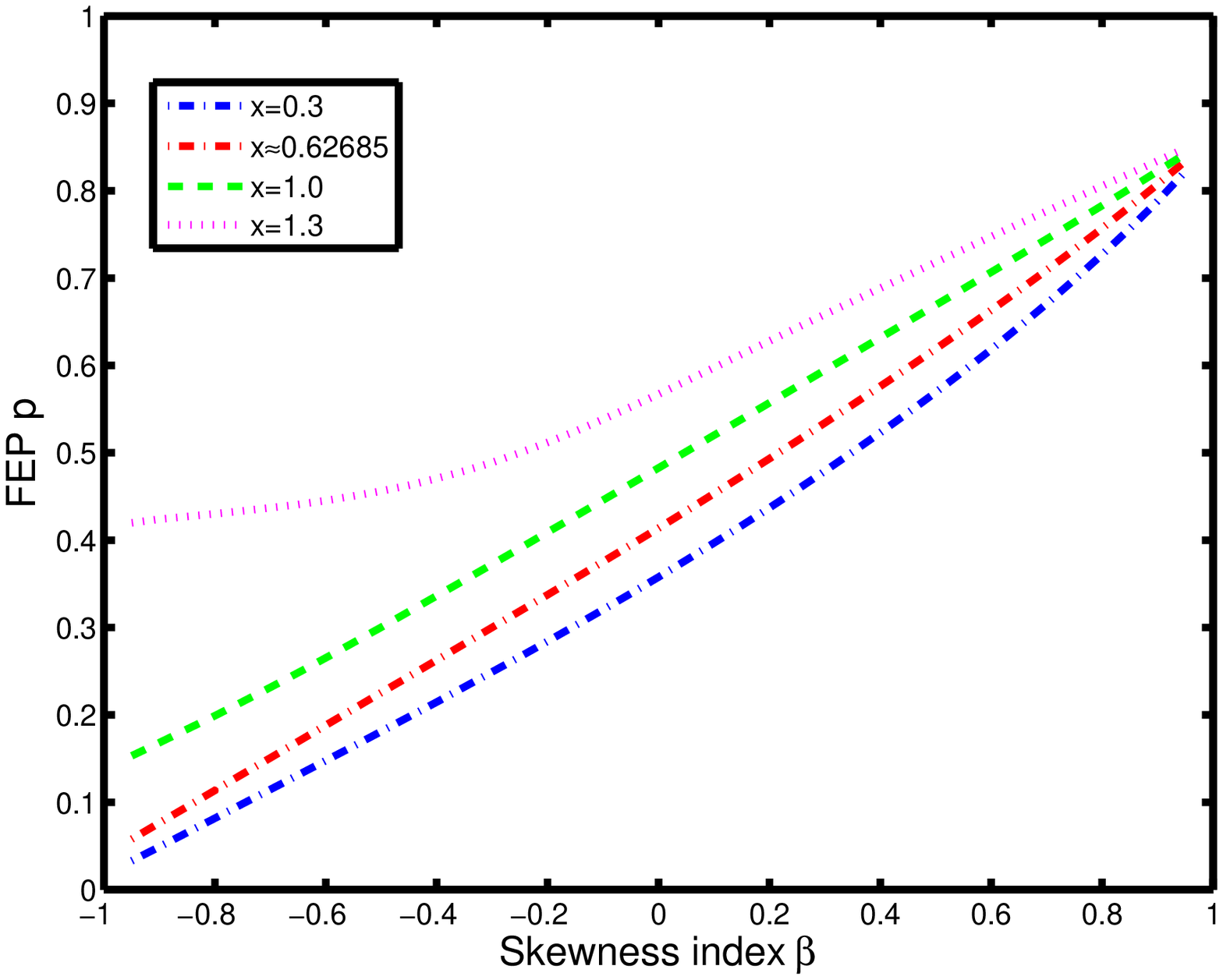}}
\subfigure[]{ \label{Fig.sub.02}
\includegraphics[width=0.45\textwidth]{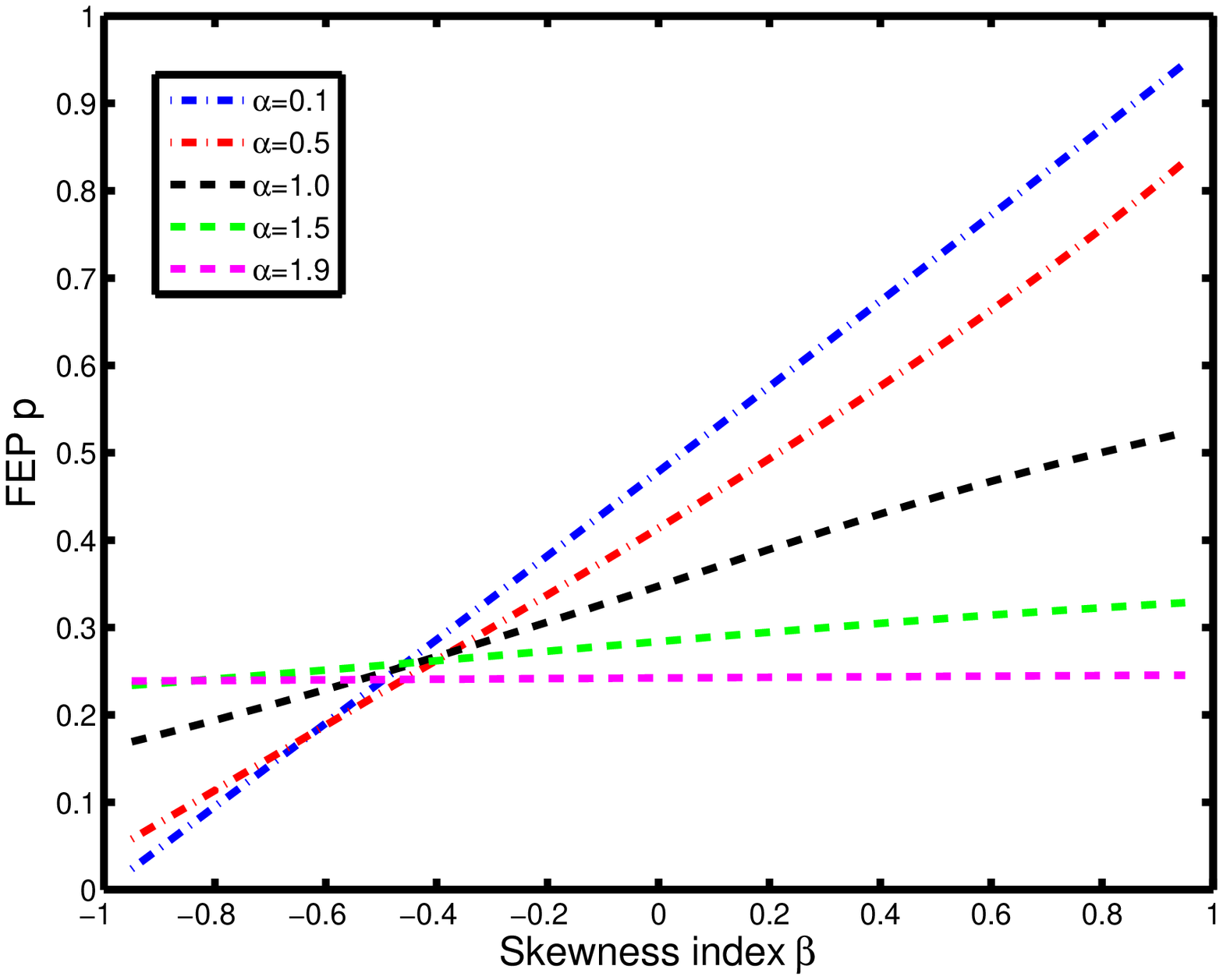}}
\caption{(Color online) FEP $p$ as a function of $\beta$,   from $D = (0, 1.48971)$ to $[1.48971,\infty)$. (a) Effect of $\beta$ and different initial concentrations  $x$ on the FEP with  $\alpha=0.5$. (b) Effect of   $\alpha$ and $\beta$  on the FEP at the lower stable concentration state $x =x_{-} \approx0.62685$.}
 \label{Fig_10}
\end{figure}

\begin{bfseries}
FEP has `turning points' with respect to $\alpha, \beta$.
\end{bfseries}
Figure \ref{Fig_10} (a) exhibits that FEP increases with the increasing $\beta$, i.e., the likelihood for transcription improves with increasing $\beta$, when the system starts in low concentrations.    When starting system at low stable concentration $x=0.62685$, we find that the evolution of FEP has  `turning points'  for   $ \beta= \beta_{turning} \approx -0.5$ (this threshold value varies slightly with various $\alpha$).  As shown in Figure \ref{Fig_10} (b),  before and after a turning   point $\beta_{turning}$,  FEP presents a  reverse relationship: Higher FEP for larger $\alpha$ suddenly switches to higher FEP for smaller $\alpha$. That is,   the higher likelihood for transcription is attained for larger non-Gaussianity index  $\alpha$ before the turning   point $\beta_{turning}$, while the opposite is true after the turning point. This phenomenon does not occur when the system is under symmetric L\'evy fluctuations.

\begin{figure}[!htb]
\subfigure[]{ \label{Fig.sub.11}
\includegraphics[width=0.45\textwidth]{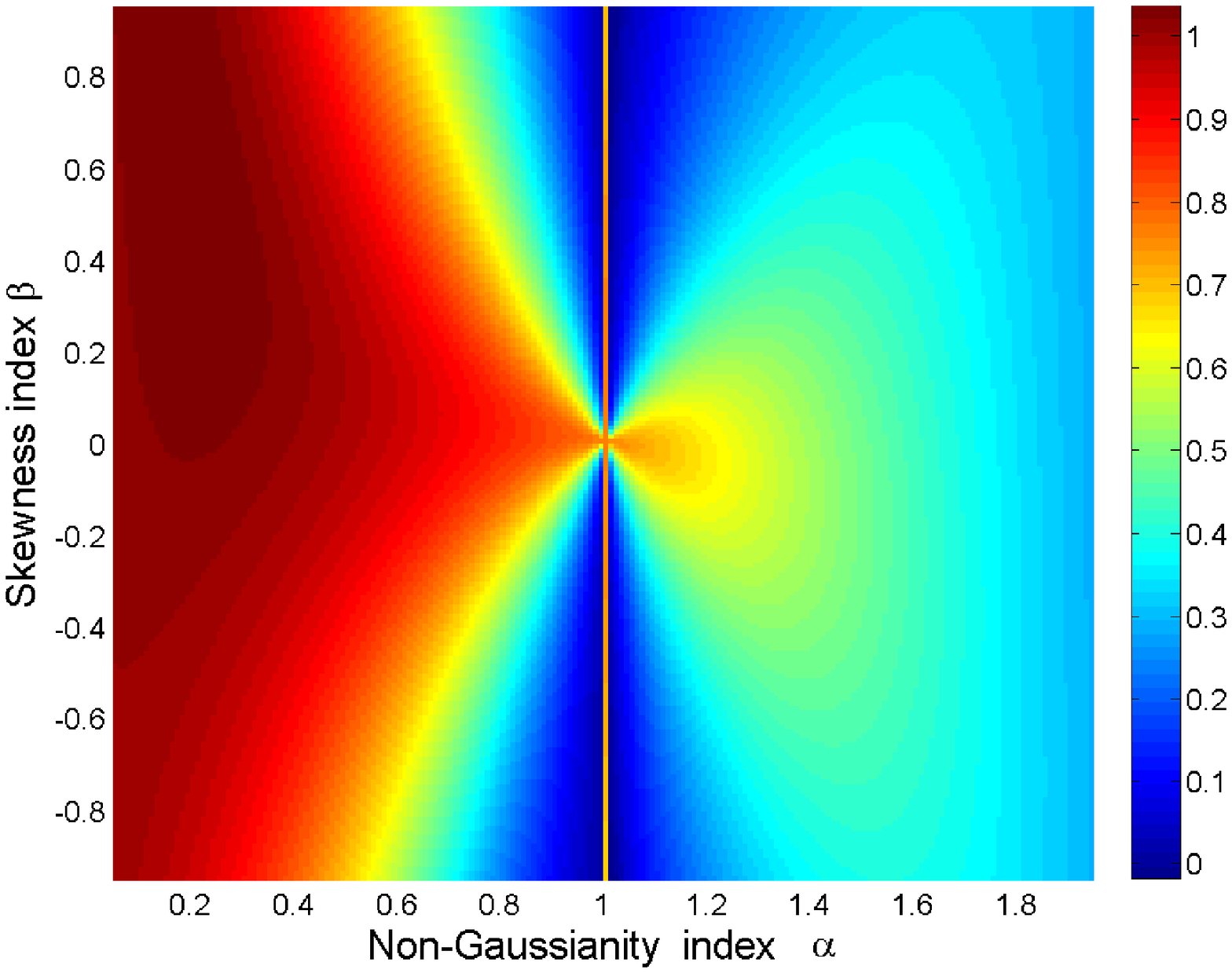}}
\subfigure[]{ \label{Fig.sub.12}
\includegraphics[width=0.45\textwidth]{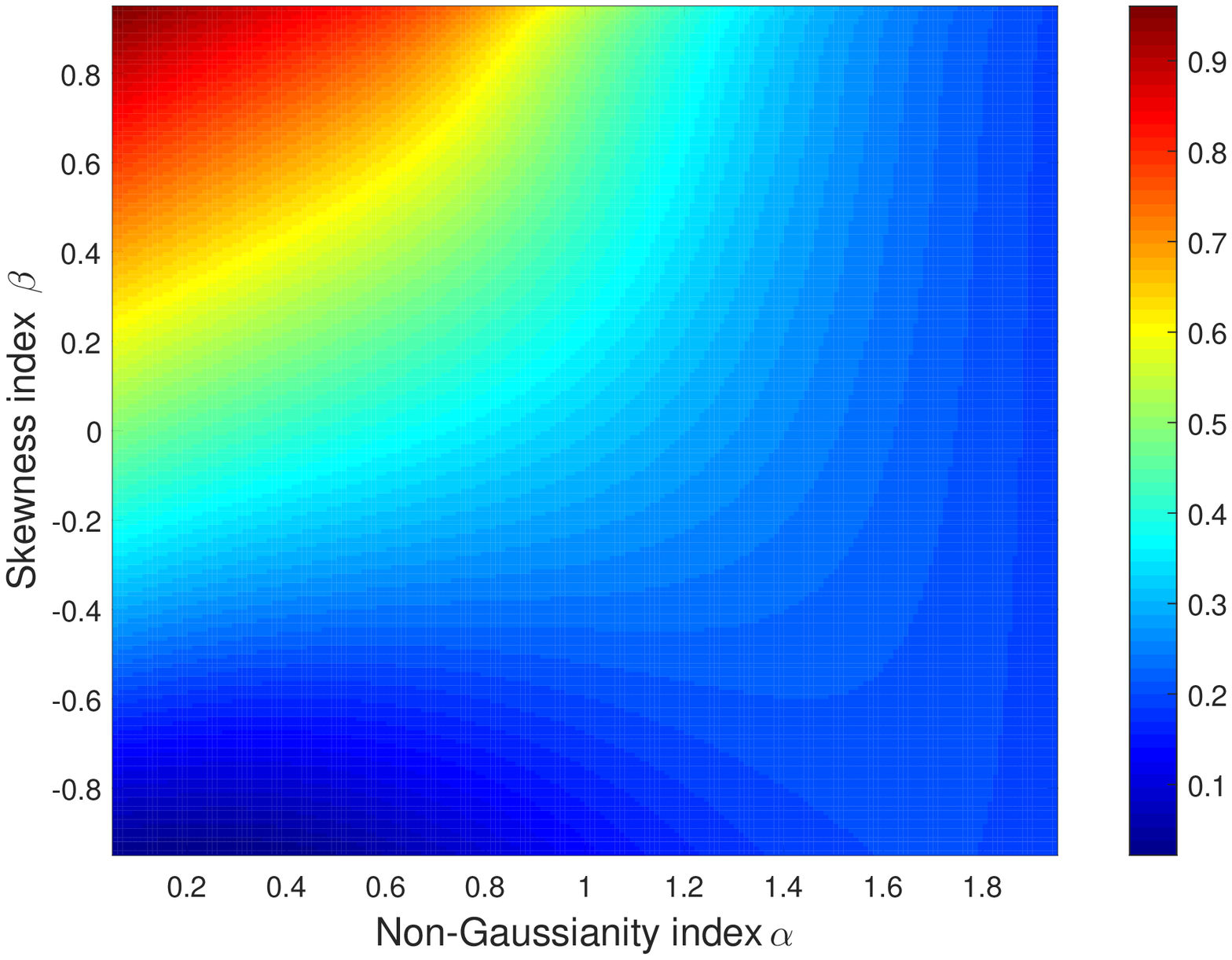}}
\caption{(Color online) (a)  $u(0.62685)$: MFET at the  lower stable  concentration state $x =x_{-} \approx 0.62685$ for   noise indexes  $(\alpha, \beta)$. (b) $p(0.62685)$:  FEP at the lower stable concentration state  $x =x_{-} \approx 0.62685$ for   noise indexes  $(\alpha, \beta)$.}
 \label{Fig_11}
\end{figure}

\begin{bfseries}
 Combined effects: Exploring the whole parameter space $\{(\alpha, \beta)\} =(0, 2) \times [-1, 1]$.
\end{bfseries}
   Figure  \ref{Fig_11} displays  the combined effects of both non-Gaussianity index $\alpha$ and  skewness index $\beta$  on MFET $u$  and FEP  $p$, at the initial concentration $x =x_{-} \approx0.62685$ (i.e., the lower stable state).

   In Figure  \ref{Fig_11}(a), the blue region indicates smaller MFET     (corresponding to  the higher likelihood for transcription), while the red region means the larger MFET. The small MFET values occur when $(\alpha, \beta)$ is in the two blue `sectors',  with $(1, 0)$ as the common vertex and with $\alpha=1$ as the separation or bifurcation line. Note that $\alpha=1$  is not a  separation point or bifurcation point in the symmetric  case $\beta=0$.  Thus, we could achieve the minimum MFET or higher likelihood for transcription by tuning the non-Gaussianity index $\alpha$ and  skewness index $\beta$ appropriately.

   In  Figure  \ref{Fig_11}(b),  we observe that the FEP is larger in the red region,  but smaller in  the blue  region.  The combined  small non-Gaussianity index $\alpha$ and  big skewness index $\beta$
  (i.e., $\alpha \in (0, 0.7) $ and $\beta \in (0.5, 1)$)  lead to bigger FEP, i.e.,  higher likelihood for transcriptions.  Therefore, we could achieve the   maximum FEP or higher likelihood for transcription by tuning the non-Gaussianity index $\alpha$ and  skewness index $\beta$ appropriately.

\section{Gene regulation with synthesis rate under multiplicative asymmetric L\'{e}vy fluctuations}
 Now we present  numerical  experiments for understanding the likelihood of transcriptions modeled by \eqref{multimodel}.

\subsection{Numerical algorithms}
The generator $A$ for stochastic differential equation model \eqref{multimodel}  is  
\begin{equation}
Au(x)= (f(x)+ \varepsilon x M_{\alpha, \beta})u'(x) +  \varepsilon \int_{\mathbb{R}^{1}\backslash \{0\}}[u(x + yx) - u(x) - I_{\{|y|<1\}}yxu'(x)]\nu_{\alpha, \beta}(dy),   \label{eq:51}
\end{equation}\\
Let $z=yg(x)$. Then the integral term  in Eq. (\ref{eq:51}) is transformed to
\begin{equation}
 g(x)+ \varepsilon x^\alpha \int_{\mathbb{R}^{1}\backslash \{0\}}[u(x + z) - u(x) - I_{\{|y|<1\}}(z)zu'(x)]\nu_{\alpha, \beta}(dz),  \label{eq:52}
\end{equation}
where
$$
g(x) =
\left \{
  \begin{array}{ll}
    \varepsilon x^\alpha u'(x)(C_2-C_1)\ln x , & \hbox{$\alpha = 1$,} \\
    \  \varepsilon x^\alpha u'(x)(C_2-C_1)\frac{x^{(1-\alpha)}-1}{1-\alpha}, & \hbox{$\alpha \neq 1$.}
  \end{array}
\right.
$$
Then we obtain that the MFET $u$ satisfies the following equation:
\begin{eqnarray}
& & c(x)u'(x)  \nonumber \\
&+&  \varepsilon x^\alpha \int_{\mathbb{R}^{1}\backslash \{0\}}[u(x+z) - u(x) - I_{\{|z|<1\}}(z)zu'(x)]\frac{[C_1 I_{\{0<z<\infty\}}(z)+C_2 I_{\{-\infty<z<0\}}(z)]}{\mid z\mid^{1+\alpha}}dz \nonumber  \\
&=& -1,     \label{eq:53}
\end{eqnarray}
where,
$$
c(x) =
\left \{
  \begin{array}{ll}
    f(x)+ \varepsilon x M_{\alpha, \beta}+ \varepsilon x^\alpha(C_2-C_1)\ln x , & \hbox{$\alpha = 1$,} \\
    \ f(x)+ \varepsilon x M_{\alpha, \beta}+\varepsilon x^\alpha (C_2-C_1)\frac{x^{(1-\alpha)}-1}{1-\alpha}, & \hbox{$\alpha \neq 1$.}
  \end{array}
\right.
$$
For  an open interval $D = (a,b)$  (in our computations below, we will take $D = (0, x_u)$),   we make a  coordinate transformation $x=\frac{b-a}{2}s + \frac{b+a}{2}$ for $s \in [-1, 1]$ and $z=\frac{b-a}{2}r$ to get finite difference discretization for  $Au(x)=-1$ as in \cite{Gao2014}:
\begin{equation}  \label{eq:55}
\begin{aligned}
(\frac{2}{b-a})c(\frac{b-a}{2}s + \frac{b+a}{2})u'(s) + \varepsilon (\frac{2}{b-a})^{\alpha} (\frac{b-a}{2}s + \frac{b+a}{2})^\alpha \int_{\mathbb{R}^{1}\backslash \{0\}}[u(s+ r) - u(s) - I_{\{|r|<1\}}(r)ru'(s)]              \\
\frac{[C_1 I_{\{0<r<\infty\}}(r)+C_2 I_{\{-\infty<r<0\}}(r)]}{\mid r \mid ^{1+\alpha}}dr=-1.
\end{aligned}
\end{equation}
With this discretization, we obtain numerical solution for nonlocal  \eqref{eq:4} and  thus MFET $u$ for stochastic model Eq. (\ref{multimodel}).
A similar method is applied to the first escape probability $p$.

\subsection{Numerical experiments}

\medskip
We summarize our major numerical simulation results below. In this section we  highlight the dynamical differences with the case of additive asymmetric L\'evy noise in the previous section.

\begin{figure}[!htb]
\subfigure[]{ \label{Fig.sub.a1}
\includegraphics[width=0.45\textwidth]{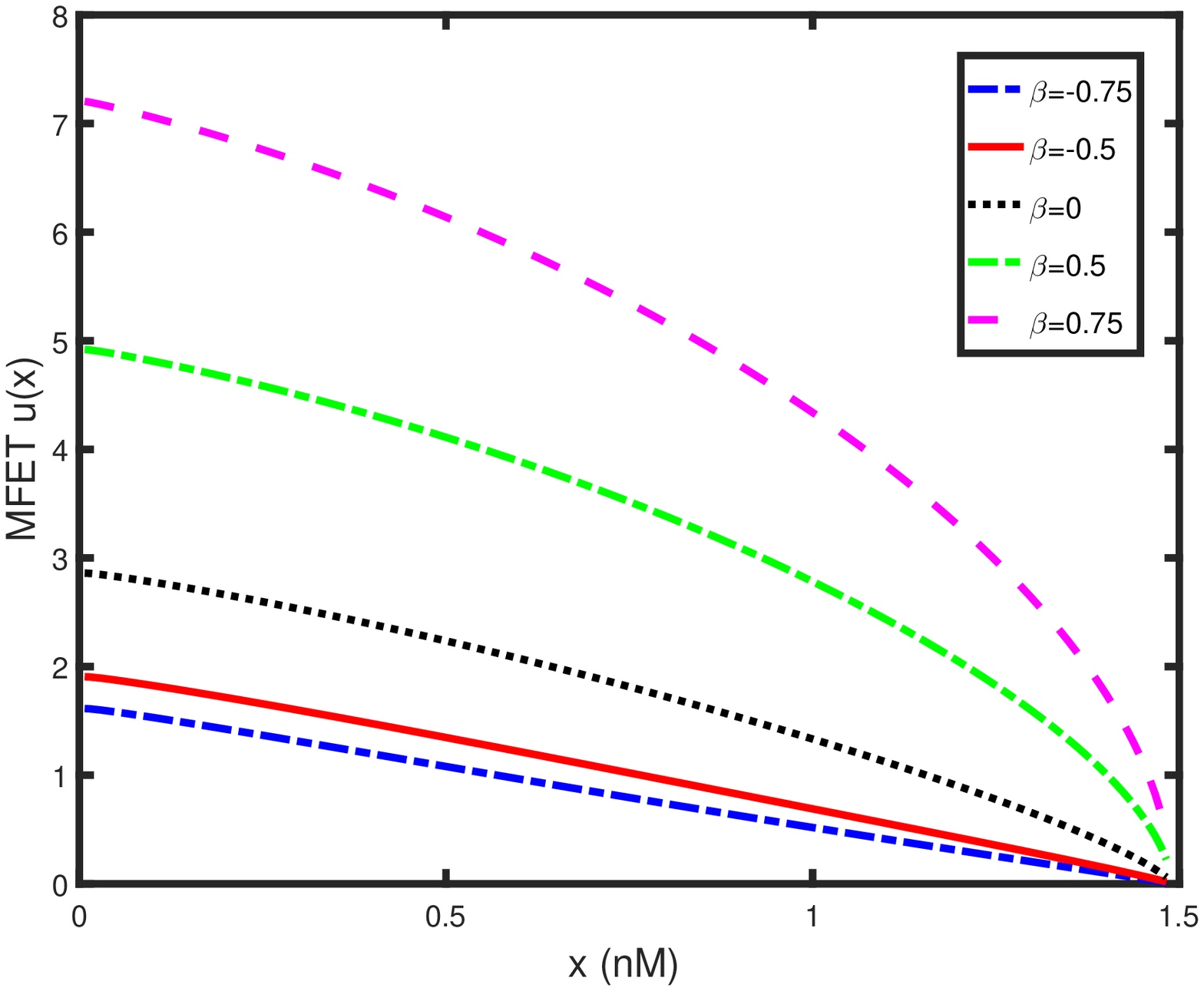}}
\subfigure[]{ \label{Fig.sub.a2}
\includegraphics[width=0.45\textwidth]{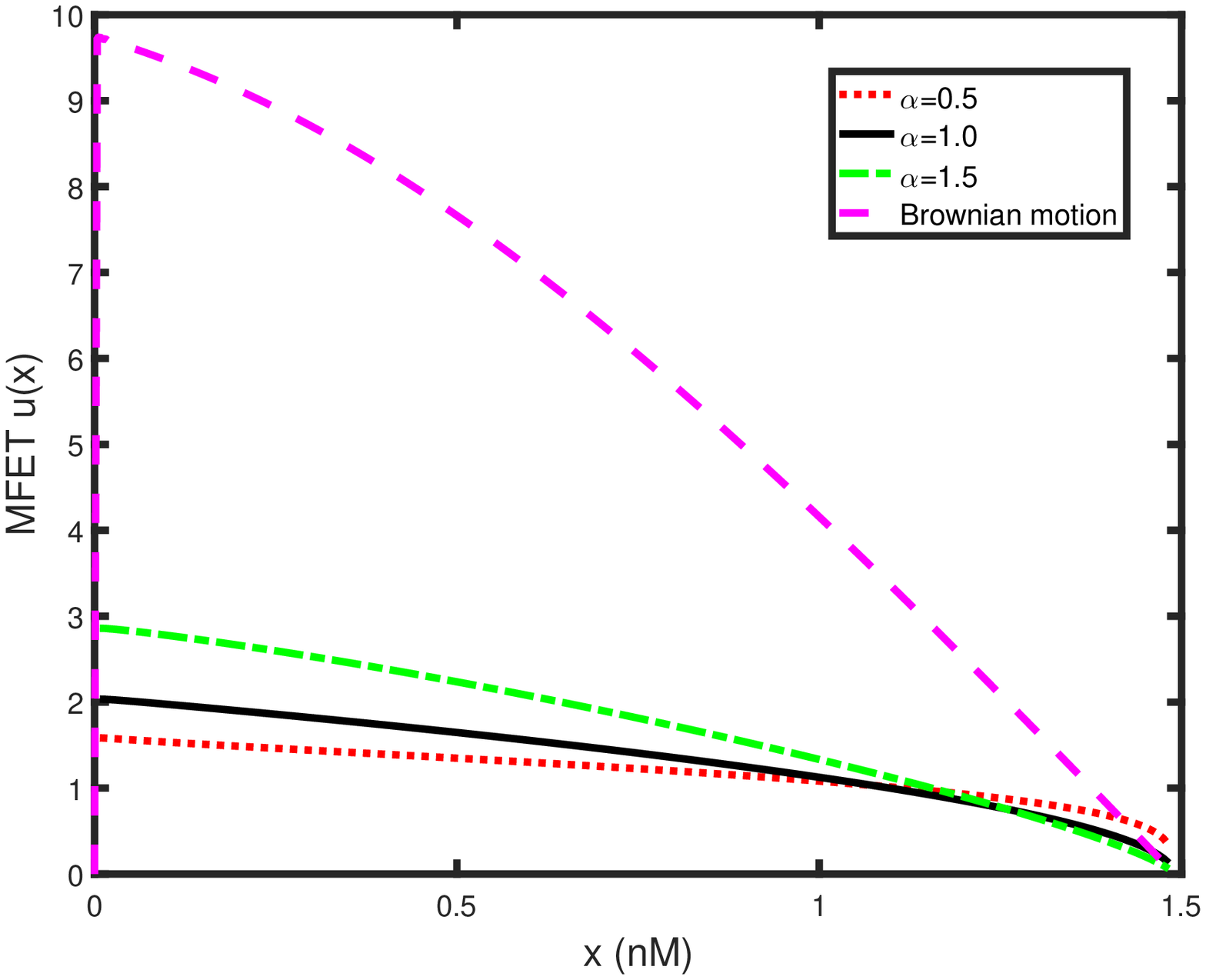}}

\caption{ (Color online) Mean first exit time (MFET)  $u(x)$ as a function of initial concentration $x$ in the low concentration domain $D = (0, 1.48971)$. (a) Effect of skewness index $\beta$ on the MFET:  $\alpha=1.5$. (b)  Effect of non-Gaussianity index  $\alpha$ on the MFET:   $\beta=0$.}
 \label{Fig_a}
\end{figure}

Figure \ref{Fig_a} plots the evolution of MFET under  multiplicative asymmetric L\'evy motion. From Figure \ref{Fig_a}(a), we observe that for $\alpha=1.5$, MFET decreases with the increase of the initial concentration $x$ in the low concentration domain. We also find that the MFET becomes short if we tune the skewness index $\beta$ small.  Figure \ref{Fig_a}(b) includes the case of multiplicative Brownian noise; in this case, the MFET is bigger than the case of multiplicative L\'evy noise clearly. MFET increases with the increase of non-Gaussianity index  $\alpha$.

\begin{figure}[!htb]
\subfigure[]{ \label{Fig.sub.b1}
\includegraphics[width=0.45\textwidth]{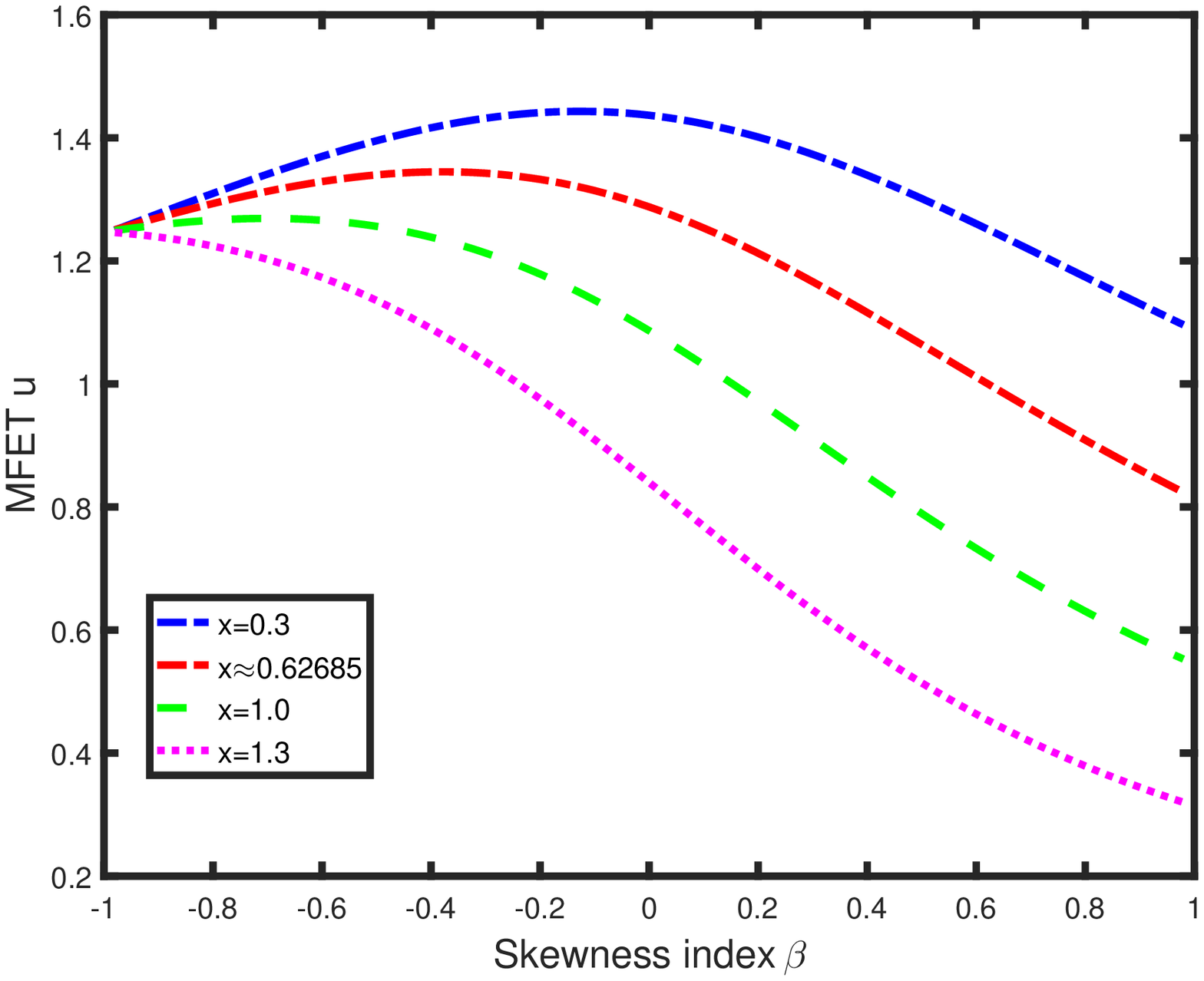}}
\subfigure[]{ \label{Fig.sub.b2}
\includegraphics[width=0.45\textwidth]{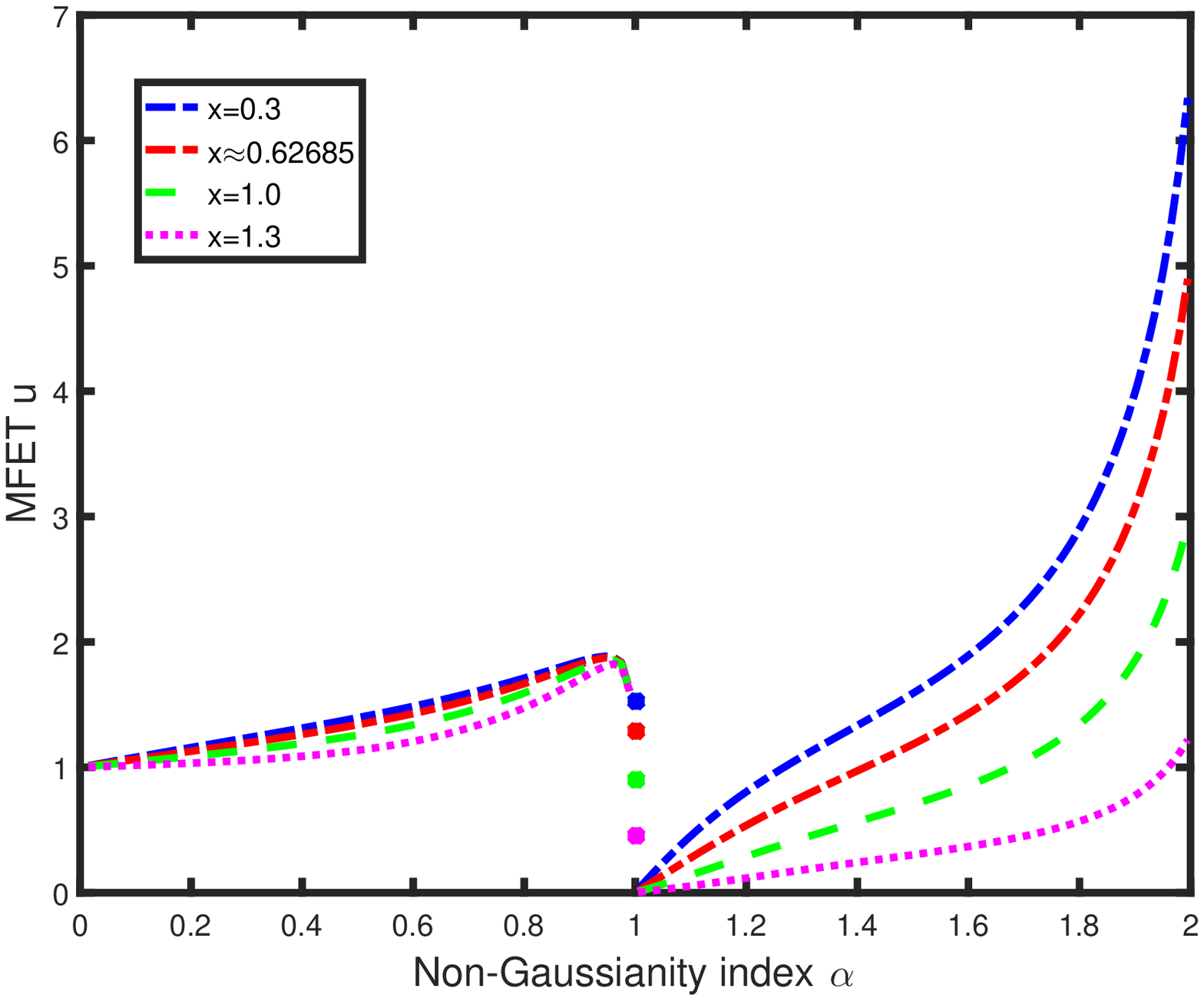}}
\caption{ (Color online) Mean first exit time (MFET) $u $ : (a) Effect of initial concentrations  $x$  and $\beta$ on the MFET:  $\alpha=0.5$.  (b) Effect of initial concentrations  $x$  and $\alpha$ on the MFET:  $\beta=-0.5$.}
 \label{Fig_b}
\end{figure}

Figure \ref{Fig_b} indicates the effects of $\alpha$ and $\beta$ on MFET for various  initial concentrations  $x$. Figure \ref{Fig_b}(a) shows the dependency in the low concentration on  the skewness index $\beta$. As is shown for the case of $\alpha=0.5$, MFET decreases with the increase of $\beta$, and the higher initial concentrations correspond to smaller MFET, i.e., higher initial concentrations $x$ benefit for the transition. Figure \ref{Fig_b}(b) exhibits that in the multiplicative asymmetric L\'evy case, MFET has a bifurcation or discontinuity point at $\alpha=1$ like in the additive asymmetric L\'evy case. For $0<\alpha<1$, MFET increases firstly then decreases when near to $\alpha=1$, however, for $1<\alpha<2$, MFET increases all the way very quickly. The maximum value is reached at $\alpha$  close to $2$.

\begin{figure}[!htb]
\subfigure[]{ \label{Fig.sub.c1}
\includegraphics[width=0.45\textwidth]{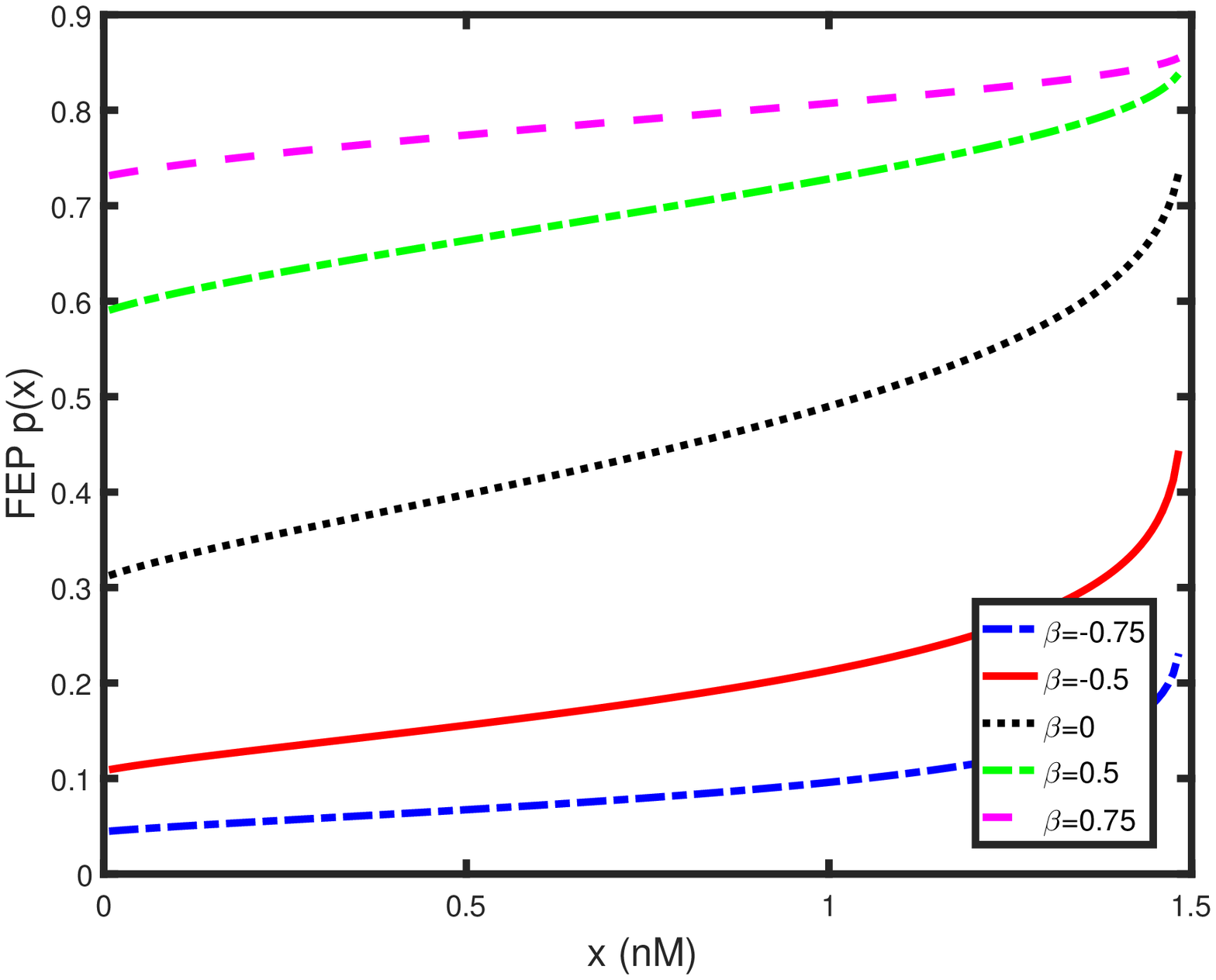}}
\subfigure[]{ \label{Fig.sub.c2}
\includegraphics[width=0.45\textwidth]{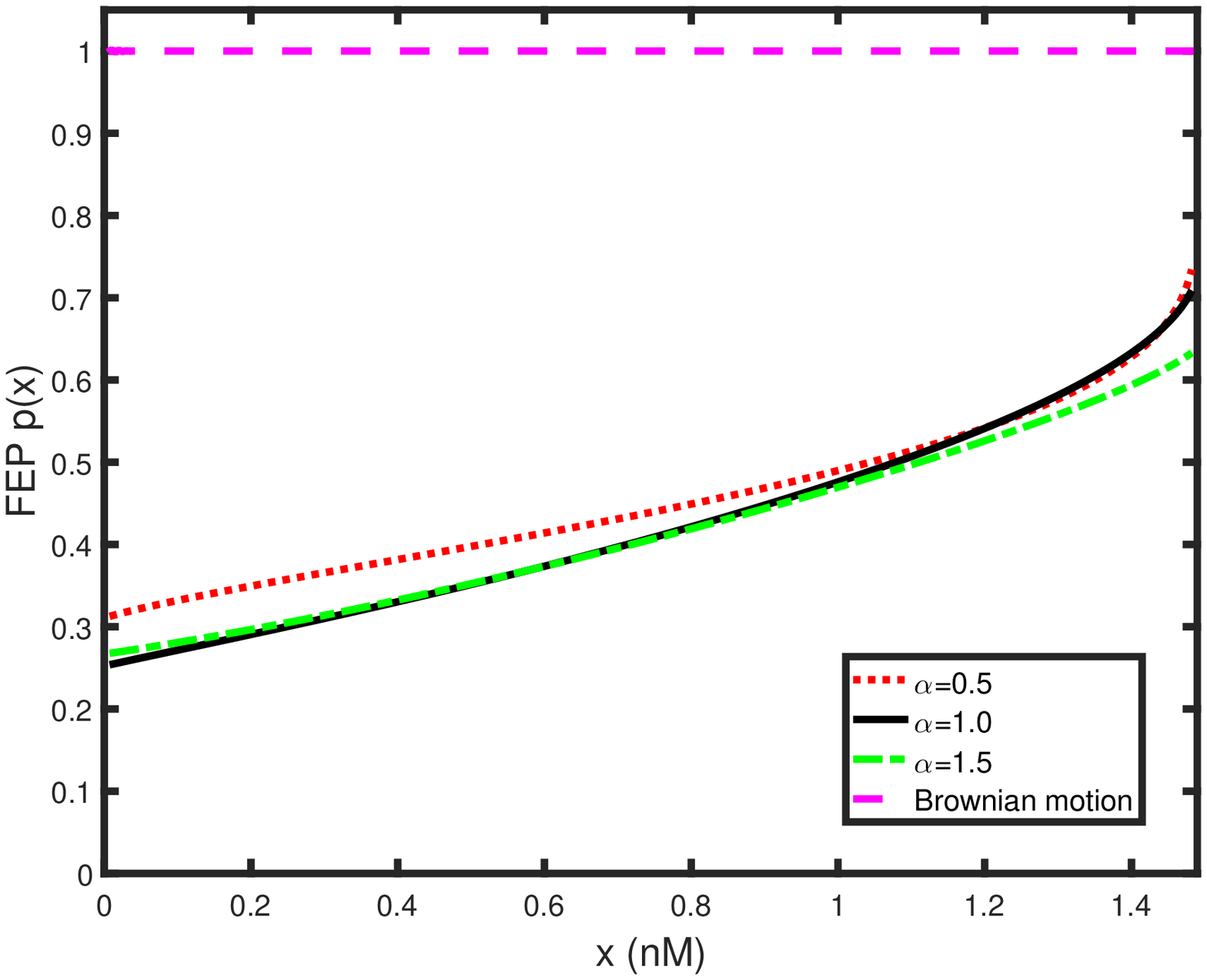}}
\caption{(Color online) FEP $p(x)$ as a function of initial concentration $x$, from $D = ( 0, 1.48971 )$ to $E=[1.48971,\infty)$.  (a) Effect of skewness index $\beta$ on the FEP : $\alpha=0.5$.  (b)Effect of non-Gaussianity index  $\alpha$ on the FEP : $\beta=0$.}
 \label{Fig_c}
\end{figure}

As shown in Figure \ref{Fig_c}(a), when $\alpha=0.5$, FEP increases with the increase of $\beta$. So the large positive $\beta$ can induce larger FEP. From Figure \ref{Fig_c}(b), we see that FEP increases in the low concentration domain with the increase of $x$, and FEP with  different $\alpha$ has intersections. Especially, the parts of the FEP for $\alpha=0.5$ overlaps with that for $\alpha=1.0$. But in the multiplicative Brownian noise case, the value of FEP is $1$, i.e., the low concentration states will get to high concentration surely, which is quite different from the additive case.

\begin{figure}[!htb]
\subfigure[]{ \label{Fig.sub.d1}
\includegraphics[width=0.45\textwidth]{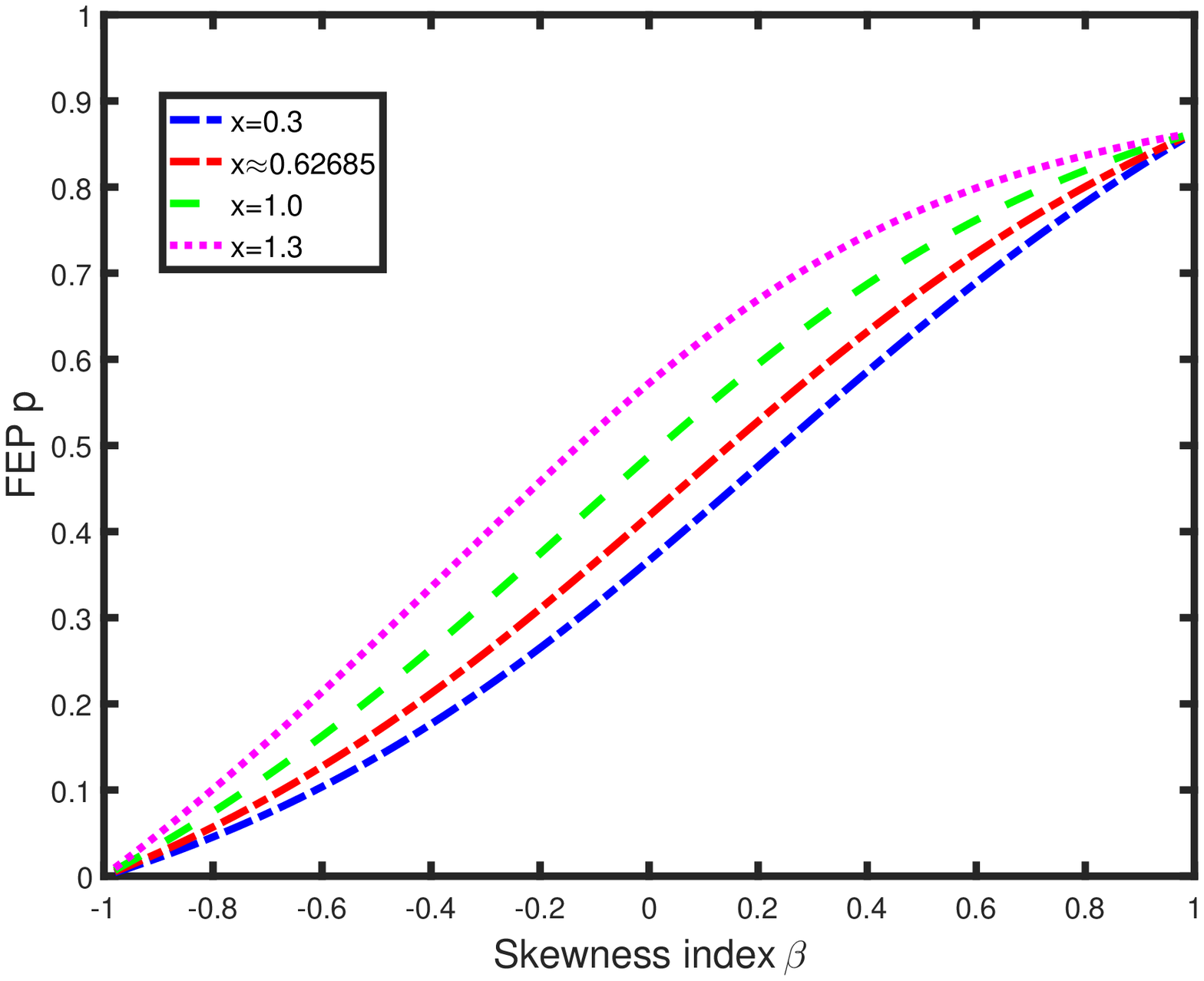}}
\subfigure[]{ \label{Fig.sub.d2}
\includegraphics[width=0.45\textwidth]{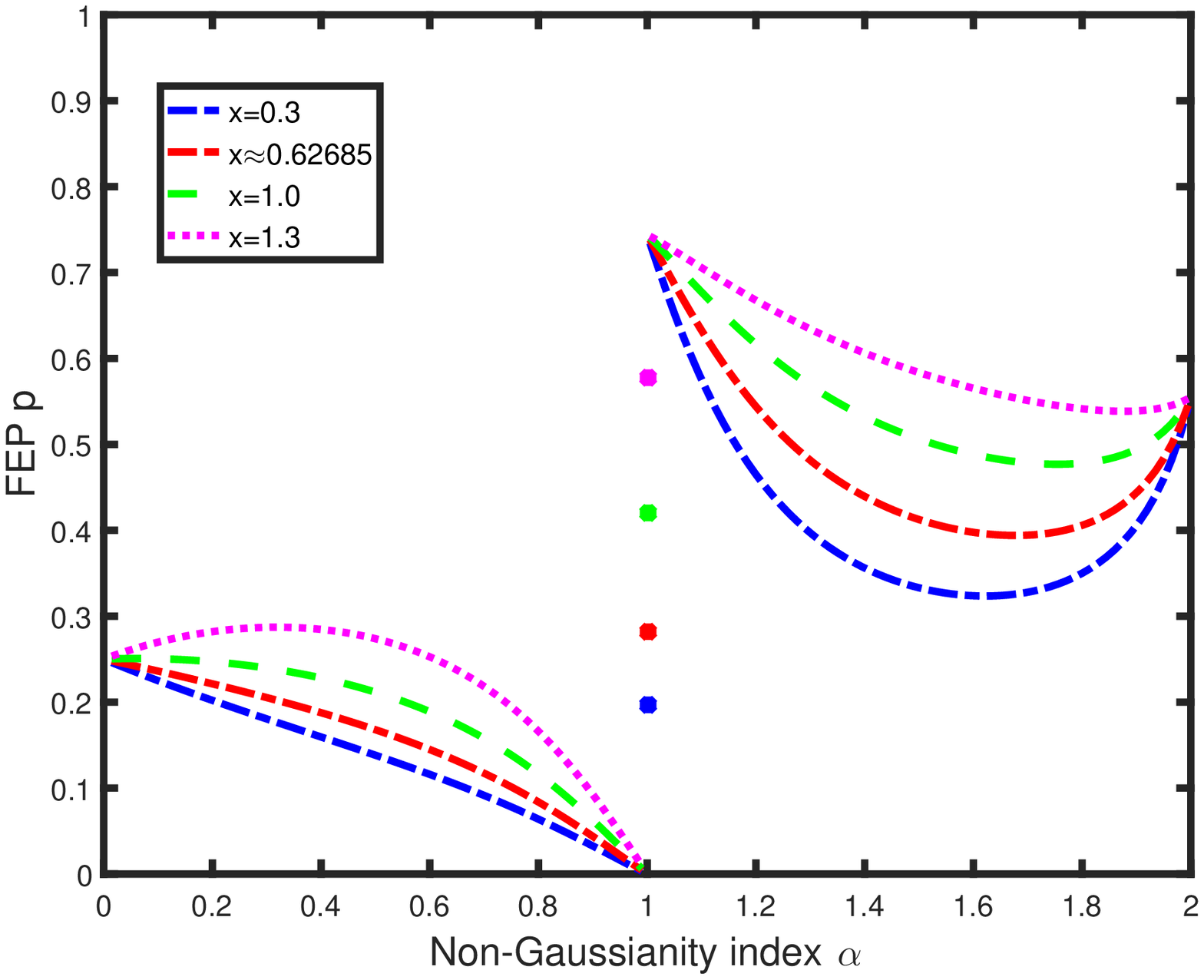}}
\caption{(Color online) First escape probability (FEP) $p$ : (a) Effect of initial concentrations  $x$  and $\beta$ on the FEP:  $\alpha=0.5$.  (b) Effect of initial concentrations  $x$  and $\alpha$ on the FEP: $\beta=-0.5$.}
 \label{Fig_d}
\end{figure}

Figure \ref{Fig_d} plots the effects of $\alpha$ and $\beta$ on FEP for various initial concentrations  $x$. Figure \ref{Fig_d}(a) shows the FEP with respect to $\beta$ for various initial concentrations  $x$. FEP increases as $\beta$ increases. Figure \ref{Fig_d}(b) shows the FEP with respect to $\alpha$ for various initial concentrations  $x$. It presents that $\alpha=1$ is also a bifurcation point for FEP. This is totally different from that in additive asymmetric L\'evy case (see Figure \ref{Fig.sub.91}). We can see that for $ 0<\alpha<1$, FEP decreases with small initial concentrations  $x$, while increases firstly then decreases with large initial concentrations  $x$. For $ 1<\alpha<2$, FEP decreases with $x$ in the low concentration domain with the increase of $\alpha$.

\begin{figure}[!htb]
\subfigure[]{ \label{Fig.sub.e1}
\includegraphics[width=0.45\textwidth]{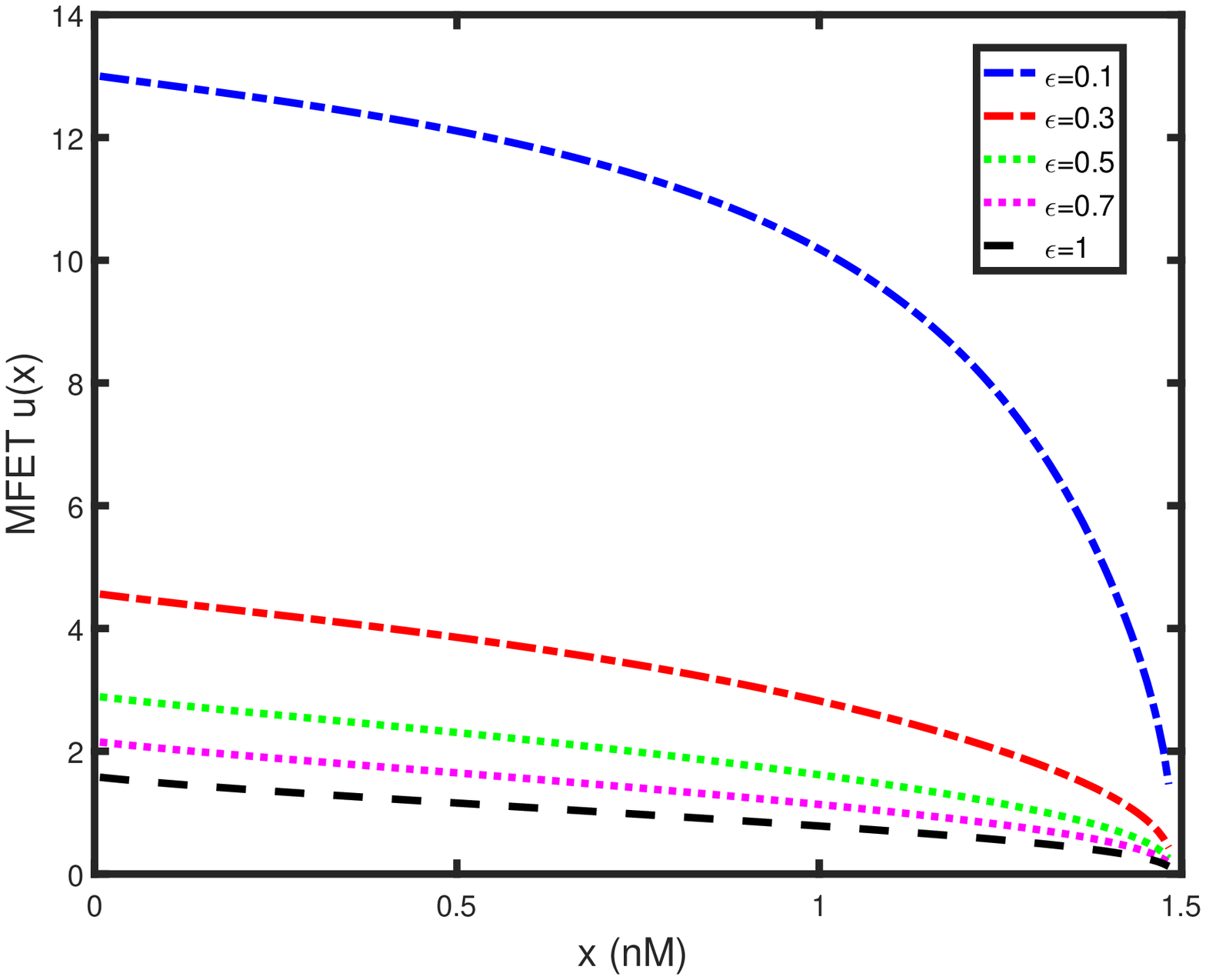}}
\subfigure[]{ \label{Fig.sub.e2}
\includegraphics[width=0.45\textwidth]{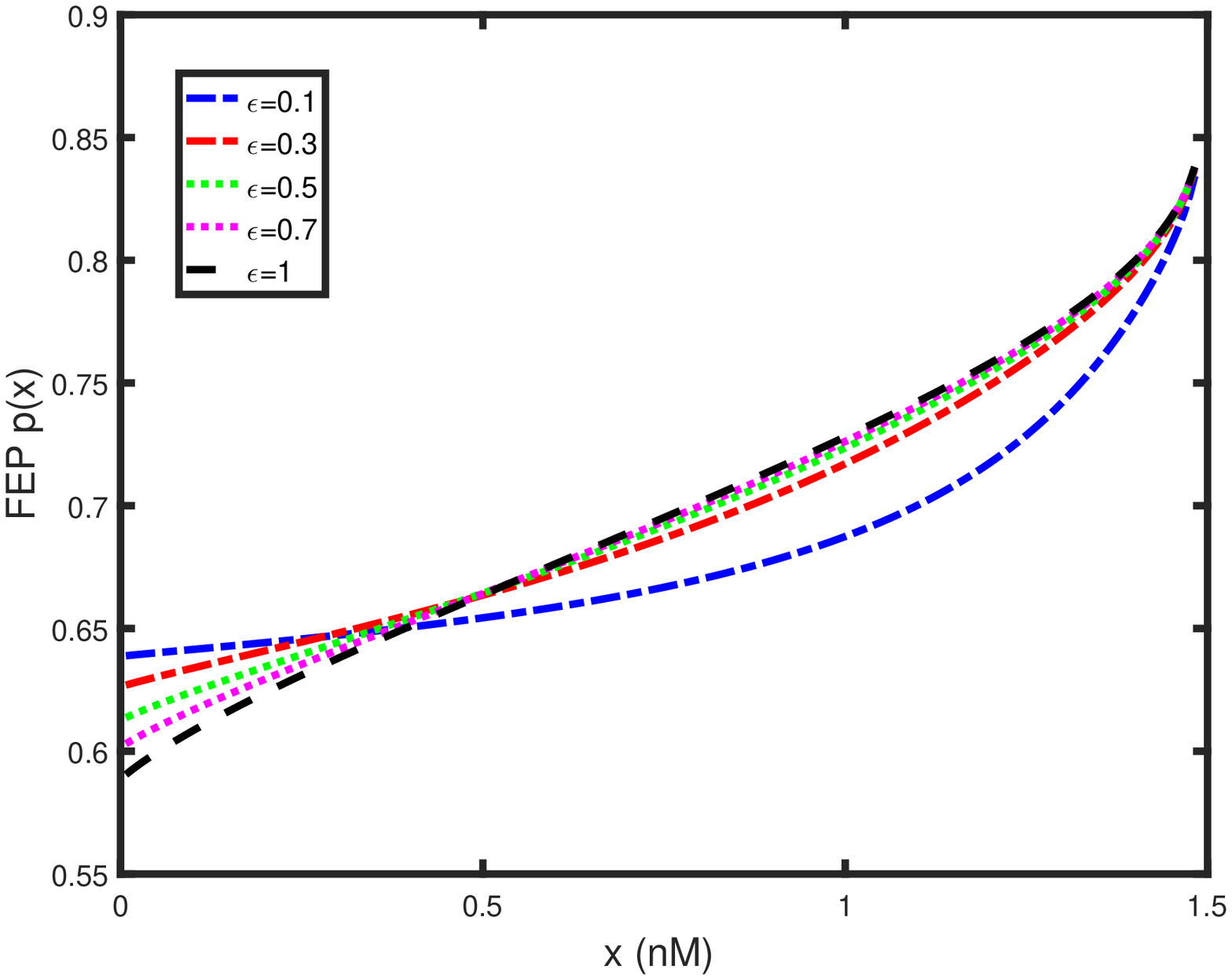}}
\caption{(Color online) (a)   MFET for various $\varepsilon$ with $\alpha=0.5, \beta=0.5$. (b) FEP for various $\varepsilon$ with $\alpha=0.5, \beta=0.5$.}
 \label{Fig_e}
\end{figure}

Figure \ref{Fig_e} demonstrates that the effects of noise  intensity $\varepsilon $ on MFET and FEP. Figure \ref{Fig_e}(a) shows us that the larger $\varepsilon $, the smaller MFET. This indicates that large noise intensity helps to exit from the low concentration domain. Figure \ref{Fig_e}(b) shows that the FEP with various  noise  intensity $\varepsilon $. The curves are crossing around $x=0.4$. Note that this phenomenon is a little complicated near the crossing point. Before and after the crossing point, FEP presents a reverse relationship with the initial concentration $x$. After the crossing point, FEP increases with the increase of $\varepsilon $.

\begin{figure}[!htb]
\subfigure[]{ \label{Fig.sub.g1}
\includegraphics[width=0.45\textwidth]{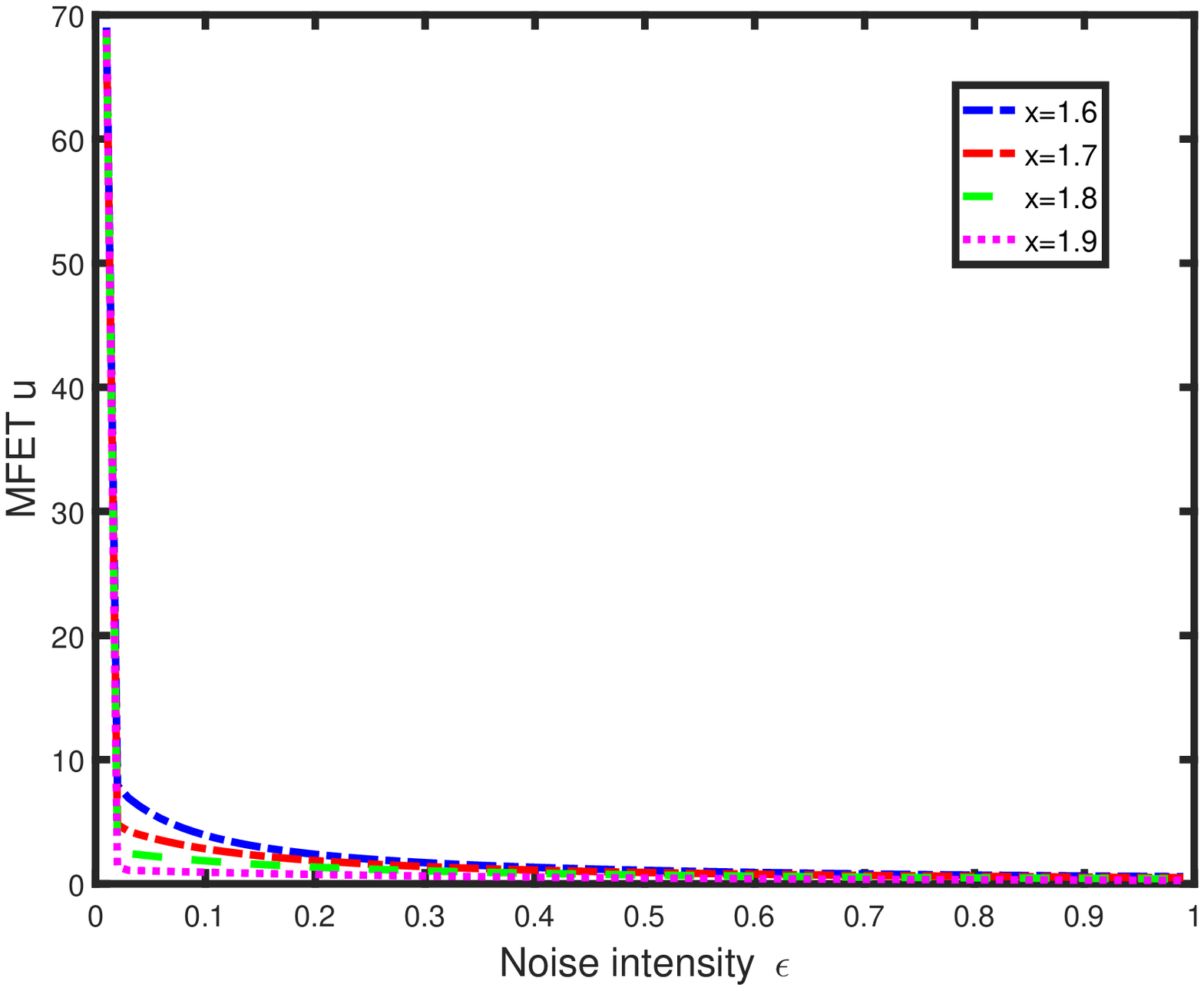}}
\subfigure[]{ \label{Fig.sub.g2}
\includegraphics[width=0.45\textwidth]{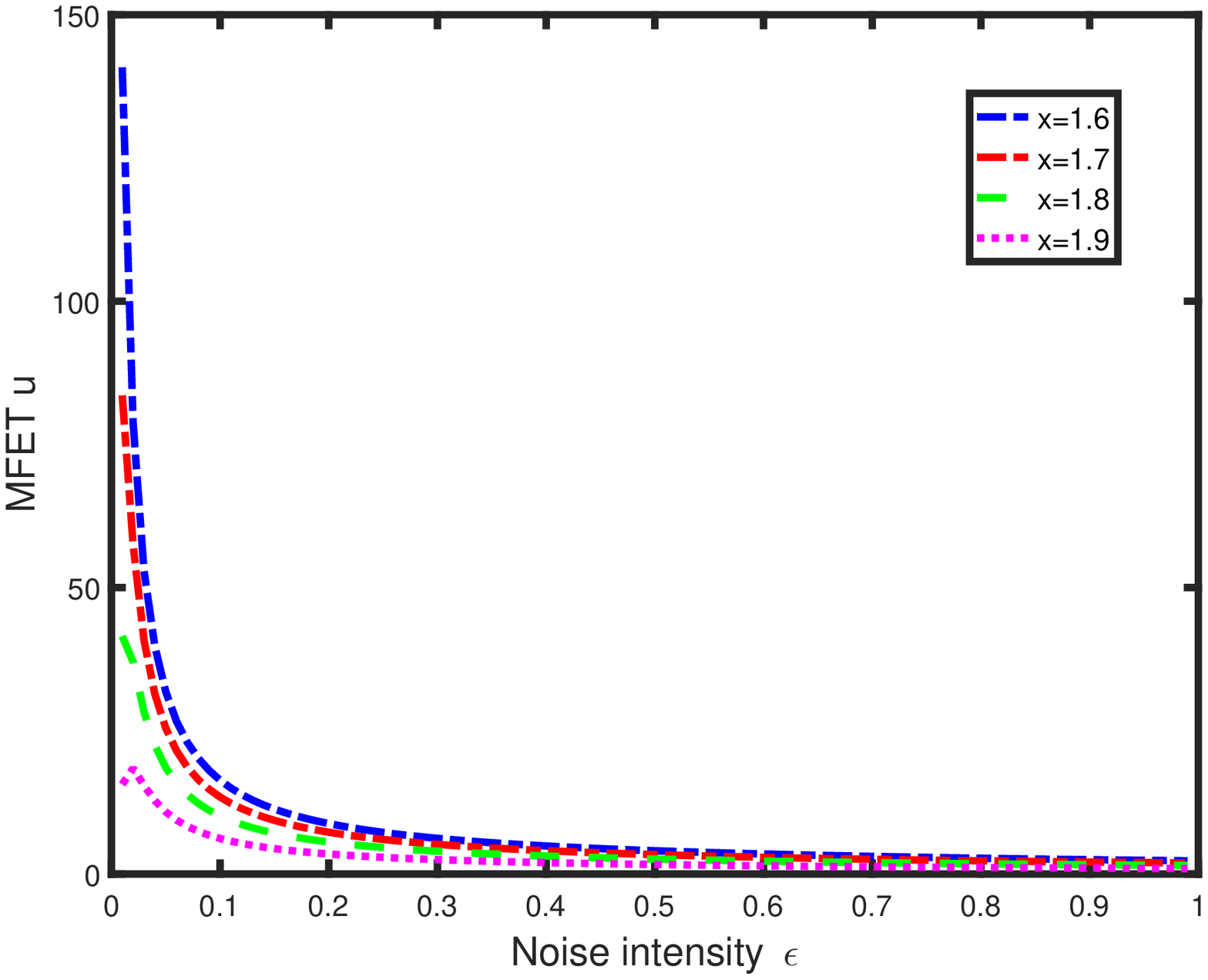}}
\caption{(Color online)   MFET as a function of $\varepsilon$ : (a) $\alpha=0.5, \beta=0.5$. (b) $\alpha=1.5, \beta=0.5$.}
 \label{Fig_g}
\end{figure}

We plot the MFET as a function of noise intensity $\varepsilon$ in Figure \ref{Fig_g}. Inspired by the literatures \cite{BS2010,BS2014,BS2016}, we are interested in computing the MFET of the stochastic genetic model \eqref{multimodel} starting from an unstable initial position, so we here set the exit domain is $[0,2]$. We could see that the MFET has a monotonic behavior with the noise intensity $\varepsilon$. The inflection point value of MFET decreases with the increase of the initial value. When MFET passes the inflection point, it changes more modestly. Comparing Figure \ref{Fig.sub.g1} and Figure \ref{Fig.sub.g2}, we could observe clearly that the smaller $\alpha$ makes the MFET shorter. The results are shown as a character of noise enhanced stability effect  \cite{BS2010,BS2014,BS2016}.

\begin{figure}[!htb]
\subfigure[]{ \label{Fig.sub.f1}
\includegraphics[width=0.45\textwidth]{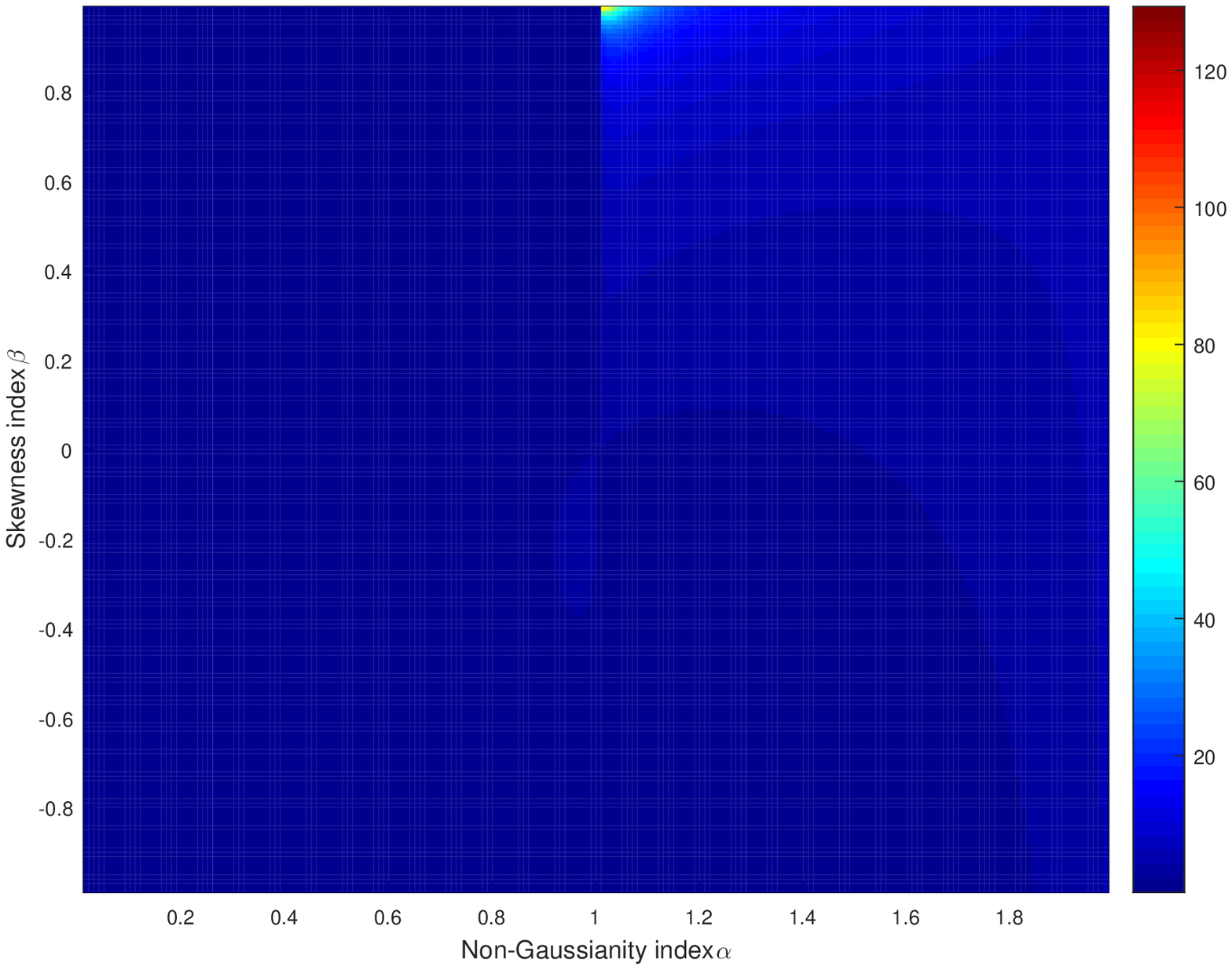}}
\subfigure[]{ \label{Fig.sub.f2}
\includegraphics[width=0.45\textwidth]{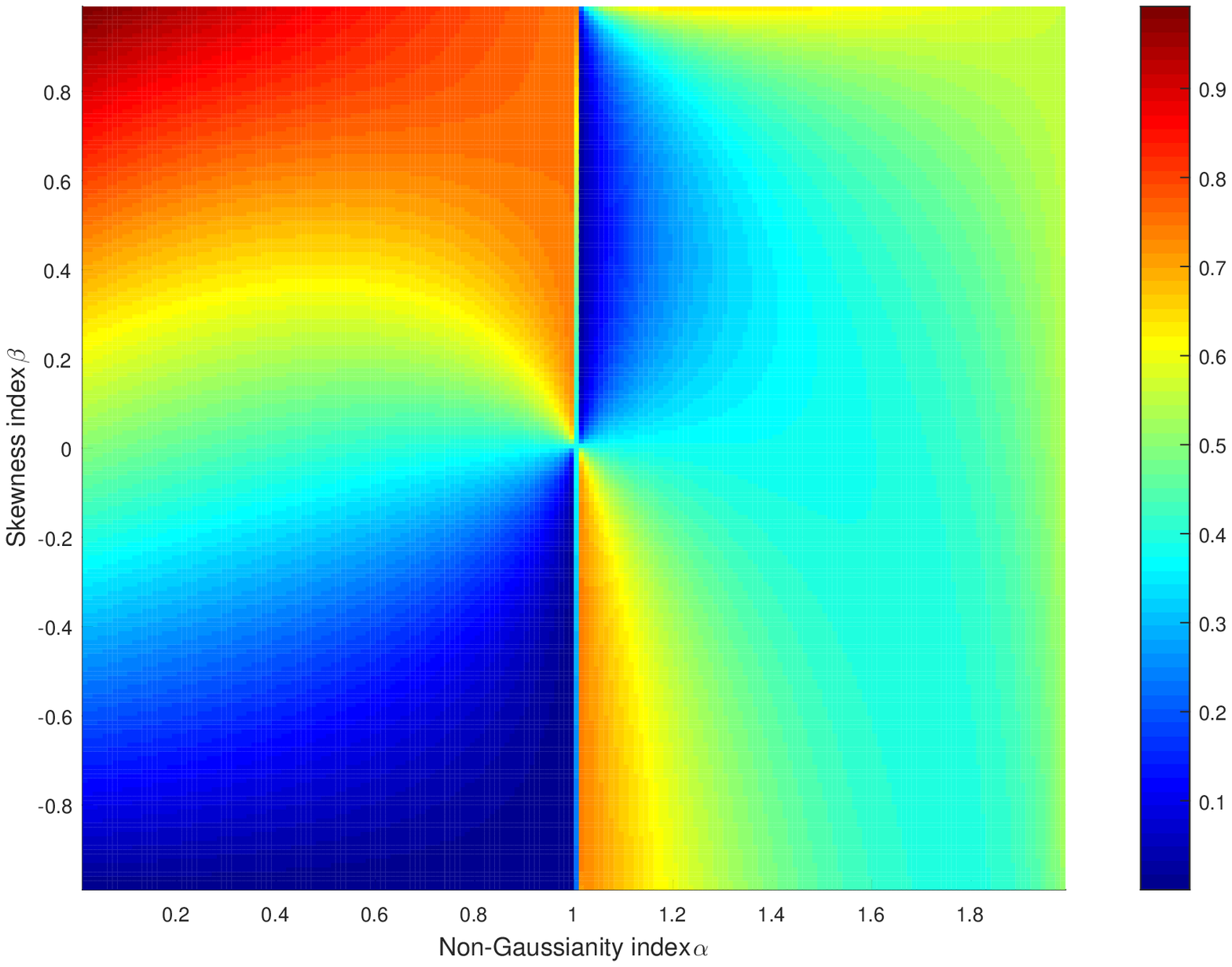}}
\caption{(Color online) (a)  $u(0.62685)$: MFET at the  lower stable  concentration state $x =x_{-} \approx 0.62685$ for   noise indexes  $(\alpha, \beta)$. (b) $p(0.62685)$:  FEP at the lower stable concentration state  $x =x_{-} \approx 0.62685$ for   noise indexes  $(\alpha, \beta)$.}
 \label{Fig_f}
\end{figure}

Figure \ref{Fig_f} presents the combined effects of non-Gaussianity index $\alpha$ and  skewness index $\beta$  on MFET $u$  and FEP  $p$, at the initial concentration $x =x_{-} \approx0.62685$ under multiplicative asymmetric L\'evy noise. From Figure \ref{Fig_f}(a), we see that smaller MFET mostly appears at $\alpha<1$. In the domain $\alpha \in (1, 1.1) $ and $\beta \in (0.9, 1)$, the MFET is rather long; we could see that $\alpha=1$ is the separation line clearly here.
In Figure \ref{Fig_f}(b), the red region presents the larger FEP domain, while the blue region represents the smaller parts. We observe that $\alpha=1$ is the  bifurcation line apparently. Note that this phenomenon does not occur in the additive asymmetric L\'evy case (see Figure \ref{Fig.sub.12}). The larger FEP values occur when $(\alpha, \beta)$ is in the two red domains. We could tune  the non-Gaussianity index $\alpha$ and  skewness index $\beta$ to the corresponding domains to achieve large FEP, i.e., having higher likelihood for transcriptions.

\section{Discussion}\label{sec5}

We have studied the effects of  asymmetric  stable L\'{e}vy noise  on a kinetic concentration model for  a genetic regulatory system. Both additive and multiplicative  asymmetric L\'{e}vy noise are considered in this paper. We have examined  possible switches or transitions from the low concentration states to the high concentration ones (i.e., likelihood for transcriptions),  excited by  asymmetric  non-Gaussian L\'{e}vy noise. Our results suggest that asymmetric  stable L\'evy noise can be used as a possible `regulator'   for gene transcriptions, for example, in the additive case,  to attain a  higher likelihood of transcription by selecting a larger positive skewness index (asymmetry index)  $\beta$  and   a small non-Gaussianity index $\alpha$. In contrast to the symmetric case, we have   observed   a bifurcation for the likelihood of transcription at the critical value $\alpha=1$ under asymmetric stable L\'evy noise ($\beta \neq 0$), as shown in Figure \ref{Fig_6}, Figure \ref{Fig_b} and Figure \ref{Fig_d}. Comparing Figure \ref{Fig.sub.12} and Figure \ref {Fig.sub.f2}, we see a striking difference between additive and multiplicative noises. There is  also a turning point in the skewness index $\beta$ for the likelihood of transcription, as seen in Figure \ref{Fig_10} (b).   The bifurcation and turning point phenomena do not occur in the symmetric noise case ($\beta = 0$).

Our results offer a possible guide to achieving certain genetic regulatory behaviors by tuning noise index \cite{Sanchez2008}, and may also provide helpful insights to further experimental research.

\paragraph{Acknowledgements:} We would like to thank Dr. Xiao Wang   for helpful discussions.

\section*{Appendix}
\renewcommand{\theequation}{A.\arabic{equation}}
\setcounter{equation}{0}


\begin{bfseries}
1. Mean first exit time
\end{bfseries}

The mean first exit time (MFET) quantifies how long the system resides  in the domain $D$ before first exit. The first  exit time is defined as follows \cite{Duan2015},
\begin{equation}
  \tau(\omega, x) = \inf\{t \geq 0: \;  X_{t}(\omega, x) \notin D\},  \;\; \omega \in \Omega,
  \label{tau}
\end{equation}
where $X_{t}(\omega, x)$ is the solution orbit  of the stochastic differential equation \eqref{eq:2}, starting with the initial \textcolor{blue}{TF-A} concentration $x$. Then the MFET is denoted as $u(x) = \mathbb{E}\tau(\omega, x)$. Here the mean $\mathbb{E}$ is taken with respect to the probability $\mathbb{P}$. The MFET $u(x)$ of the solution orbit  $X_{t}(\omega, x)$, starting with the initial \textcolor{blue}{TF-A} concentration $x$,  is the mean time to stay in the low concentration domain $D$.    \\
Denote the generator of  the  stochastic differential equation  \eqref{eq:2} by $A$.
It  is defined as $Au = \lim_{t \rightarrow 0}\frac{P_t u - u}{t}$, where $P_{t}u(x) = \mathbb{E}u(X_t)$. The generator  $A$ for the gene regulatory system \eqref{eq:2} is explicitly  given in  \eqref{eq:7},  Section 4.1.
Then the  mean exit time $u$ satisfies the following nonlocal equation \cite{Duan2015} with an exterior boundary condition
\begin{eqnarray} \label{eq:4}
   Au(x) &=& -1, \qquad  x \in D,\notag\\
  u(x)  &=& 0, \qquad  x \in D^{c}.
\end{eqnarray}
Here $ D^{c}$ is the complement set of $D$ in $\mathbb{R}^{1}$.

When we take the domain $D = (0, x_u)$, containing the low concentration stable state ``$x_-$",  the  MFET is the mean time scale for the system to exit  the low concentration state. The longer the mean exit time is, the less likely the system is in transcription.

\begin{bfseries}
2. First escape probability
\end{bfseries}

The first escape probability (FEP), denoted by $p(x)$,  is the likelihood that   the \textcolor{blue}{TF-A}  monomer,  with initial concentration $x$, first escapes from the low  concentration domain $D$ and lands in the high concentration domain $E$. That is,
\begin{equation} \label{eq:5}
  p(x) = \mathbb{P}\{X_\tau(x) \in E\},
\end{equation}
where $\tau$ is the exit time from $D$, as in  \eqref{tau}.
 This  first escape probability $p$ satisfies the following nonlocal equation \cite{Duan2015} with a special, exterior boundary   condition:
\begin{eqnarray} \label{eq:6}
  Ap(x) &=& 0,   \;  x\in D,  \notag\\
  p(x) &=& 1,   \; x\in E,  \notag\\
  p(x) &=& 0,  \;  x \in D^c \setminus E,
\end{eqnarray}
where $A$ is the  generator for the stochastic differential equation   \eqref{eq:2}, as in  \eqref{eq:7},  Section 4.1.







\end{document}